\documentclass[a4paper,11pt]{article}

\usepackage{ulem}
\usepackage{graphicx}
\usepackage{tikz}
\usepackage{tikz-cd}
\usetikzlibrary{decorations.pathmorphing}

\usepackage{jheppub} 

\usepackage{ytableau}
\usepackage{esint}
\usepackage[table]{xcolor}
\definecolor{nosaka}{rgb}{0.0, 0.5, 0.0}

\title{\boldmath New recursion relations for M2-brane matrix models}

\author[a,c]{Bin He}
\emailAdd{bhe22@m.fudan.edu.cn}
\affiliation[a]{
Shanghai Institute for Mathematics and Interdisciplinary Sciences,\\
Block A, International Innovation Plaza, No.~657 Songhu Road, Yangpu District, Shanghai, China
}

\author[a,b]{and Tomoki Nosaka}
\emailAdd{nosaka@simis.cn}

\affiliation[b]{
Center for Mathematics and Interdisciplinary Sciences, Fudan University, Shanghai, 200433, China
}

\affiliation[c]{
 Department of Physics and Center for Field Theory and Particle Physics, Fudan University, 2005 Songhu Road, Shanghai 200438, China
 }

\abstract{
In this paper we investigate the finite $N$ exact values of the $S^3$ partition function of the ${\cal N}=4$ super Yang-Mills theory with one adjoint hypermultiplet and $N_\text{f}$ fundamental hypermultiplets, which describes $N$ M2-branes on $\mathbb{C}^2\times \mathbb{C}^2/\mathbb{Z}_{N_\text{f}}$, with mass and FI deformations.
We claim that the grand canonical sum of the partition function obeys a bilinear difference relation with respect to the shifts of the mass parameters of the fundamental hypermultiplets, which results in a new recursion relation for the partition function with respect to $N$.
As an application, we also determine the analytic expression for the leading $1/N$ non-perturbative correction to the free energy of these models, which would correspond holographically to the contribution from an M2-brane wrapped on a 3d volume in the internal space of $\text{AdS}_4\times S^7/\mathbb{Z}_{N_\text{f}}$.
}

\allowdisplaybreaks
\begin{document}
\maketitle

\ytableausetup{boxsize=1.5mm}

\flushbottom

\section{Introduction and Summary}

The 3d superseymmetric gauge theories describing multiple M2-branes in M-theory are important objects both in theoretical physics and mathematics.
A representative example of such theories is the ${\cal N}=6$ $\text{U}(N)_k\times \text{U}(N+M)_{-k}$ superconformal Chern-Simons matter theory called ABJ theory \cite{Aharony:2010af,Hosomichi:2008jd,Aharony:2008gk}.
This theory consists of an ${\cal N}=2$ $\text{U}(N)$ Chern-Simons vector multiplet with the Chern-Simons level $k$, an ${\cal N}=2$ $\text{U}(N+M)$ Chern-Simons vector multiplet with the Chern-Simons level $-k$, a bifundamental hypermultiplets and a bifundamental twisted hypermultiplets, and describes a stack of $N$ M2-branes probing $\mathbb{C}^4/\mathbb{Z}_k$ with $M$ units of discrete torsion, which can be regarded as $M$ ``fractional M2-branes'' localized at the orbifold singularity.
Generalizations of this theory with various different gauge groups or quiver diagrams were also proposed \cite{Hosomichi:2008jd,Aharony:2008gk,Hosomichi:2008jb,Imamura:2008dt,Gulotta:2012yd}, which describe M2-branes probing different background geometries.

In particular, the $S^3$ partition function of these theories, which are expressed as finite dimensional ordinary integrations due to the method of supersymmetric localization \cite{Kapustin:2009kz} and hence we call them ``M2-brane matrix models'', exhibit various interesting properties.
First of all, the free energy, which is the logarithm of the partition function, exhibits $N^{3/2}$ scaling in the large $N$ limit.
This behavior, together with the prefactor of $N^{3/2}$, is consistent with the free energy of the eleven dimensional supergravity on $\text{AdS}_4$ dual to the supersymmetric gauge theory under the AdS/CFT correspondence \cite{Klebanov:1996ag,Maldacena:1997re}.
Furthermore, a special class of M2-brane matrix models including that of ABJ theory can be rewritten in the Fermi gas formalism \cite{Marino:2011eh}, which allows us to determine the all order $1/N$ perturbative corrections as the expansion coefficients of the Airy function governed only by three model-dependent parameters.
The Fermi gas formalism relates the matrix model to the spectral problem of a one-dimensional quantum mechanical system, which can be viewed as the quantum version of a complex one-dimensional curve determined by the inverse of the Fermi gas density matrix $\hat\rho$.
It was proposed \cite{Grassi:2014zfa,Codesido:2015dia} that, when the one-dimensional curve is the mirror curve of a toric Calabi-Yau threefold, the quantization condition of the spectral problem is exactly determined by the free energy of the refined topological string on the Calabi-Yau threefold.
This proposal is called topological string/spectral theory (TS/ST) correspondence and was confirmed through various different setups \cite{Hatsuda:2013oxa,Kallen:2013qla,Honda:2014npa,Huang:2014eha,Kallen:2014lsa,Wang:2014ega,Marino:2015ixa,Hatsuda:2015oaa,Kashaev:2015wia,Codesido:2015dia,Okuyama:2015pzt,Grassi:2016vkw,Codesido:2016ixn,Moriyama:2017gye,Zakany:2017txl,Codesido:2017jwp}.
This can also be rephrased that the grand canonical partition function of the M2-brane matrix model with the Fermi gas formalism is given by the partition functions of the refined topological string, which completely determines the $1/N$ expansion of the M2-brane matrix model including the non-perturbative effects in $1/N$.
Such exact results on the $1/N$ corrections provide non-trivial prediction to the quantum effects/higher derivative corrections in the dual gravity setup, some of which were also reproduced from direct analyses of the gravity side \cite{Bhattacharyya:2012ye,Beccaria:2023ujc}.\footnote{
See also \cite{Dabholkar:2014wpa,Caputa:2018asc,Cassia:2025aus,Gautason:2025plx,Cassia:2025jkr} for the attempts to explain the all order $1/N$ corrections from the gravity side.
}

At the same time, based on the plenty of the exact values of the ABJ matrix model for finite $k,N,M$, it was found that the grand partition function of the ABJ matrix model satisfy non-linear difference relations with respect to the relative rank $M$ \cite{Grassi:2014uua}.
These relations can also be reorganized into the Hirota bilinear form the $\mathfrak{q}$-deformed Painlev\'e $\text{III}_3$ equation \cite{Bonelli:2017gdk}.
The existance of such bilinear equation for the grand partition function is also suggested from the TS/ST correspondence.
Under the TS/ST correspondence, the grand partition function of the ABJ matrix model is related to the topological string partition function on local $\mathbb{P}^1\times \mathbb{P}^1$.
Through the geometric engineering this is further identified with the partition function of 5d ${\cal N}=1$ $\text{SU}(2)$ pure Yang-Mills theory on $S^1\times \mathbb{C}^2$ with $\Omega$ deformation.
This 5d partition function is known to satisfy the $\mathfrak{q}$-Painlev\'e $\text{III}_3$ equations \cite{Bershtein:2016aef} in the context of the $K$-theoretic uplift of the Painlev\'e/gauge theory correspondence \cite{Bonelli:2016qwg} (or the isomonodromy/CFT correspondence \cite{Gamayun:2012ma} combined with the AGT relation \cite{Alday:2009aq}), which can also be derived directly from the Nakajima-Yoshioka blowup equations \cite{MR2183121,Bershtein:2018zcz}.
The blowup equations are non-linear relations for the 5d partition function obtained by evaluating the partition function on $S^1$ times the blowup of $\mathbb{C}^2$ at the origin, and the similar relations are expected to hold for more general 5d theories.
This suggests that the grand partition function of a more general M2-brane matrix model written in the Fermi gas formalism would also satisfy some non-linear relations analogous to the $\mathfrak{q}$-Painlev\'e $\text{III}_3$ equations in the ABJ matrix model.
This was indeed confirmed for a four-node $\text{U}(N_1)_k\times \text{U}(N_2)_0\times \text{U}(N_3)_{-k}\times \text{U}(N_4)_0$ circular quiver Chern-Simons theory \cite{Bonelli:2022dse,Moriyama:2023mjx}, whose the grand partition satisfy the bilinear equations of $\mathfrak{q}$-Painlev\'e VI system \cite{Jimbo:2017ael,Stoyan:2025qya}.

On the other hand it was found that the bilinear equation for the ABJ matrix model also extends \cite{Nosaka:2020tyv,Nosaka:2024gle} to the ABJ matrix model with two-parameter real mass deformations \cite{Hosomichi:2008jb,Gomis:2008vc,Jafferis:2011zi}.
Although this model enjoys the Fermi gas formalism, its density matrix has no clear identification with a 5d theory for generic complex values of the mass parameters.
This may suggest that there is a more direct way to understand the bilinear relations from the matrix models or 3d supersymmetric gauge theories, rather than through the TS/ST correspondence.
Finding such new connection would also be useful for deepening our understanding of the TS/ST correspondence.
In addition, as was noticed in \cite{Moriyama:2023pxd,Nosaka:2024gle}, the bilinear equations combined with the 3d Seiberg-like dualities can result in a simple recursion relation for the matrix models with respect to $N$.
Such recursion relation allows us to determine the exact values of the matrix models for finite but very large $N$, and hence it can be a new powerful tool to study the large $N$ expansion of the M2-brane matrix models.
In \cite{Nosaka:2024gle} the recursion relation indeed played crucial role in order to reveal the large $N$ behavior of the partition function of the mass deformed ABJ theory when the mass parameters are larger than the critical values, where the Airy function no longer reproduces the actual behavior of the partition function \cite{Honda:2018pqa}.

The purpose of this paper is to deepen our understanding of the bilinear equations satisfied by the M2-brane matrix models.
So far the bilinear equations are know explicitly only for the (mass deformed) ABJ theory and the four-node quiver Chern-Simons theory mentioned above.
Therefore, it would be useful to first find out more examples of the M2-brane matrix models satisfying bilinear equations.
For this purpose, we first recall the bilinear equation for the (mass deformed) $\text{U}(N)_k\times \text{U}(N+M)_{-k}$ ABJ theory.
Originally the bilinear equations were found as the relation among the grand partition functions with different values of $M$.
When $k=1$, however, by using the dualities induced by the Hanany-Witten effect \cite{Hanany:1996ie} in the type IIB brane setup of the ABJ theory, we can rewrite the bilinear equation purely in terms of the grand partition function of $\text{U}(N)_1\times \text{U}(N)_{-1}$ ABJM theory.
Since the ABJM theory with $k=1$ is dual to the 3d ${\cal N}=4$ $\text{U}(N)$ super Yang-Mills theory with one adjoint hypermultiplet and one fundamental hypermultiplet, the bilinear equation for the mass deformed ABJ theory can also be viewed as the bilinear equation for this super Yang-Mills theory.

In this paper we consider a generalization of this setup to the super Yang-Mills theory with $N_\text{f}$ hypermultiplets with $N_\text{f}\ge 2$.
When the rank $N$ is not too large, the finite $N$ exact values of the partition function of these models can be obtained systematically by finite residue sums.
By using these exact values, we can find the bilinear equations for the grand partition functions of these models which are satisfied at least for a first few order in the fugacity dual to $N$.
As in the mass deformed ABJ theory, these bilinear equations can be rewritten into conjectural recursion relations for the partition functions with respect to $N$.
We find that the exact values of the partition function for finite but large $N$ generated by these recursion relations show an excellent agreement with the Airy form obtained by the Fermi gas formalism of these models \cite{Kubo:2025dot}.
This result strongly suggests that the bilinear equations are satisfied for all order in the fugacity.

In addition, based on the residue sum approach, we also propose a simple combinatorial formula for the partition function for general rank $N$ and the flavors $N_\text{f}$.
Although the calculation of residues involve a subtlety for higher ranks, the final formula is also verified by the recursion relations obtained from the bilinear equations.
Our combinatorial formula is a non-trivial extension of the formula obtained in \cite{Gaiotto:2019mmf} for $N_\text{f}=1$.

As an application of the recursion relation, we also investigate the leading $1/N$ non-perturbative corrections to the free energy of the super Yang-Mills theory.
When the mass and FI parameters are small, the gravity dual of our setup would be a smooth deformation of $\text{AdS}_4\times S^7/\mathbb{Z}_{N_\text{f}}$ (see e.g.~\cite{Freedman:2013oja,Gautason:2023igo}), where $S^7/\mathbb{Z}_{N_\text{f}}$ is a radial section of $\mathbb{C}^2\times \mathbb{C}^2/\mathbb{Z}_{N_\text{f}}$.
The $1/N$ non-perturbative corrections would correspond on the gravity side to the contributions from the M2-branes wrapped on the three-dimensional volumes in the internal space of this geometry.

The rest of this paper is organized as follows.
In section \ref{sec_themodel} we fix our convention for the $S^3$ partition function of the ${\cal N}=4$ $\text{U}(N)$ super Yang-Mills theory with one adjoint  hypermultiplet and $N_\text{f}$ fundamental hypermultiplets, which we shall call ``$N_\text{f}$ matrix model''.
In section \ref{sec_N123} we evaluate $N_\text{f}$ matrix models for $N=1,2,3$ by finite residue sums.
In section \ref{sec_generalcombinatorial} based on these results and the combinatorial formula for $N_\text{f}=1$ \cite{Gaiotto:2019mmf} we propose the combinatorial formula for $N_\text{f}$ matrix model for general $N_\text{f}$.
In section \ref{sec_bilineq} we investigate the bilinear equations for the $N_\text{f}$ matrix models based on the exact values for $N=1$ and $N=2$.
We also rewrite them into the recursion relations and provide non-trivial checks for the combinatorial formula proposed in section \ref{sec_generalcombinatorial}.
In section \ref{sec_1stwscoef} we use the recursion relations to determine the analytic expression for the coefficient of the leading non-perturbative effect in $1/N$.
In section \ref{sec_Discuss} we summarize our results and discuss possible directions for further investigations.
In appendix \ref{app_Youngdiagramnotation} we explain our convention for the Young diagrams used in section \ref{sec_generalcombinatorial}.
In appendix \ref{app_proofofgtildevsNekrasov} we explain how to simplify a couple of redundant factors in the combinatorial formula \eqref{gtilde} into the product of fewer factors \eqref{Nf2combinatorialformula_Elambdamuagain}.

\section{The model}
\label{sec_themodel}

In this paper we consider the 3d ${\cal N}=4$ $\text{U}(N)$ Yang-Mills theory with one adjoint hypermultiplet and $N_\text{f}$ fundamental hypermultiplets.
The partition function of this theory on $S^3$ can be calculated by the supersymmetric localization formula \cite{Kapustin:2009kz}, and we obtain the following formula
\begin{align}
&Z_{y_\alpha}^{(N_\text{f})}(\zeta,m;N) = \nonumber \\
&\frac{1}{\left(2\cosh\frac{m}{2}\right)^N}\frac{1}{N!}\int_{-\infty}^\infty \frac{d^Nx}{(2\pi)^N}\,e^{-\frac{i\zeta}{2\pi}\sum_{i=1}^Nx_i}\frac{\prod_{i<j}^N\left(2\sinh\frac{x_i-x_j}{2}\right)^2}{\prod_{i<j}^N\prod_\pm 2\cosh\frac{x_i-x_j\pm m}{2} \prod_{i=1}^N\prod_{\alpha=1}^{N_\text{f}}2\cosh\frac{x_i-y_\alpha}{2}}.
\label{ADHMgeneralNf}
\end{align}
Here $\zeta$ is the Fayet-Illiopoulos parameter, $m$ is the mass parameter of the adjoint hypermultiplet and $y_\alpha$ are the mass parameters of the fundamental hypermultiplets.
Without loss of generality, we impose the condition
\begin{align}
\sum_{\alpha=1}^{N_\text{f}}y_\alpha =0
\end{align}
to $y_\alpha$.
In this paper we shall refer this partition function to $N_\text{f}$ matrix model.

In \cite{Nosaka:2015iiw,Kubo:2025dot} it was found by using the Fermi gas formalism that the large $N$ expansion of the $N_\text{f}$ matrix model is given by the Airy form
\begin{align}
Z^{(N_\text{f})}_{y_\alpha}(\zeta,m;N)=Z^{(N_\text{f})}_{\text{pert},y_\alpha}(\zeta,m;N)\left(1+e^{-\# \sqrt{\frac{N-B}{C}}}\right),
\label{ZNfpertAiry}
\end{align}
with
\begin{align}
Z_{\text{pert},y_\alpha}^{(N_\text{f})}(\zeta,m;N)=e^{A}\,C^{-\frac{1}{3}}\text{Ai}\left[C^{-\frac{1}{3}}(N-B)\right].
\label{ZNfpertAiry2}
\end{align}
Here the parameters $C,B,A$ are given by \cite{Kubo:2025dot}\footnote{
The $N_\text{f}$ matrix model \eqref{ADHMgeneralNf} is equivalent to the $S^3$ partition function of $\text{U}(N)_k\times \text{U}(N)_0^{\otimes (N_\text{f}-1)}\times \text{U}(N)_{-k}$ super Chern-Simons matter theory with $k=1$ considered in section 2 of \cite{Kubo:2025dot}, where the parameters in \eqref{ADHMgeneralNf} are related to the parameters in \cite{Kubo:2025dot} as
\begin{align}
q^{\text{[KNP]}}=N_\text{f},\quad
M^{\text{[KNP]}}=\frac{\zeta}{\pi N_\text{f}},\quad
\eta_\alpha^{\text{[KNP]}}=\frac{y_\alpha}{\pi},\quad
\tilde{q}^{\text{[KNP]}}=1,\quad
\tilde{M}^{\text{[KNP]}}=\frac{m}{\pi},\quad
\tilde{\eta}_\alpha^{\text{[KNP]}}=0,\quad
k^{\text{[KNP]}}=1.\nonumber
\end{align}
}
\begin{subequations}
\label{CBA}
\begin{align}
&C=\frac{2}{\pi^2 N_\text{f}\left(1+\frac{\zeta^2}{\pi^2 N_\text{f}^2}\right)\left(1+\frac{m^2}{\pi^2}\right)},\label{CofCBA}\\
&B=-\frac{1}{6N_\text{f}\left(1+\frac{\zeta^2}{\pi^2 N_\text{f}^2}\right)}-\frac{1}{2\left(1+\frac{m^2}{\pi^2}\right)}\left(\sum_{\alpha=1}^{N_\text{f}}\frac{y_\alpha^2}{\pi^2}+\frac{N_\text{f}}{3}\right)
+\frac{2}{3N_\text{f}\left(1+\frac{\zeta^2}{\pi^2 N_\text{f}^2}\right)\left(1+\frac{m^2}{\pi^2}\right)}
+\frac{N_\text{f}}{24},\\
&A=\frac{1}{4}\sum_\pm\biggl({\cal A}\Bigl(N_\text{f}\pm\frac{i\zeta}{\pi},0\Bigr)
+\sum_{\alpha,\beta=1}^{N_\text{f}}{\cal A}\Bigl(1\pm\frac{im}{\pi},\frac{y_\alpha-y_\beta}{\pi}\Bigr)\biggr),
\end{align}
\end{subequations}
with
\begin{align}
\mathcal{A}(\kappa,\chi)=\frac{2\zeta(3)}{\pi^2\kappa}+\frac{\chi^2}{2\kappa}-\frac{\kappa}{12}+\frac{1}{\pi}\int_0^\infty dz\,\frac{1}{e^{2\pi z}-1}\frac{d}{dz}\biggl(\frac{\cos \pi\chi z}{z\tanh\frac{\pi\kappa z}{2}}-\frac{2}{\pi\kappa z^2}\biggr).
\end{align}
Note that for $\chi=0$ the function ${\cal A}(\kappa,\chi)$ reduces to the integration which gives the parameter $A$ of the Airy form for the $\text{U}(N)_\kappa\times \text{U}(N)_{-\kappa}$ ABJM theory \cite{Hatsuda:2014vsa}
\begin{align}
{\cal A}(\kappa,0)=
\frac{2\zeta(3)}{\pi^2\kappa}\Bigl(1-\frac{\kappa^3}{16}\Bigr)+\frac{\kappa^2}{\pi^2}\int_0^\infty
dz\,\frac{z}{e^{\kappa z}-1}\log(1-e^{-2z}).
\end{align}
In this paper we mainly focus on the $N_\text{f}$ matrix model \eqref{ADHMgeneralNf} at finite $N$ rather than their large $N$ expansion.
Nevertheless, the Airy form provides a useful check for the validity of the finite $N$ exact values of the partition function.

In the following sections, we often omit the arguments $\zeta,m$ and denote the partition function $Z^{(N_\text{f})}_{y_\alpha}(\zeta,m;N)$ simply as $Z^{(N_\text{f})}_{y_\alpha}(N)$ in the absence of confusion.

\section{Exact partition function by finite residue sum}
\label{sec_N123}

As investigated in \cite{Gaiotto:2019mmf} for $N_\text{f}=1$, the $N_\text{f}$ matrix model \eqref{ADHMgeneralNf} can be evaluated efficiently by the residue sum.
This method is also applicable for $N_\text{f}\ge 2$, which goes as follows.
For later purpose, let us define
\begin{align}
R_m=e^{\frac{i\zeta m}{2\pi}},\quad
R_{y_\alpha}=e^{\frac{i\zeta y_\alpha}{2\pi}}.
\label{RmRy}
\end{align}

\subsection{$N=1$}

\begin{figure}
\begin{center}
\begin{tikzpicture}[scale=0.3]
\draw[->] (-5,0)--(25,0);
\draw[->] (10,0)--(10,10);
\draw[color=red] (0,0.05)--(20,0.05)--(20,6)--(0,6)--(0,0.05);
\draw[color=red,->] (14,0.05)--(16,0.05);
\draw (22,8)--(25,8);
\draw (22,8)--(22,10);
\node [above] at (10,6) {$2\pi i$};
\node [below] at (20,0) {$\infty$};
\node [below] at (0,0) {$-\infty$};
\node [above] at (23.5,8.5) {$x_i$};
\end{tikzpicture}
\caption{The contour $\gamma$ used in the integration \eqref{ZN1withcountourgamma}.}
\label{boxcontour}
\end{center}
\end{figure}
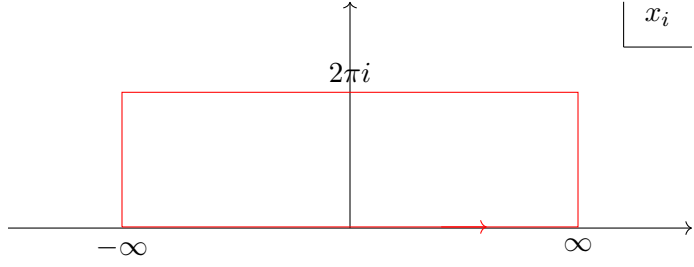

First let us consider the case with $N=1$
\begin{align}
Z_{y_\alpha}^{(N_\text{f})}(1)=\frac{1}{2\cosh\frac{m}{2}}\int_{-\infty}^\infty\frac{dx_1}{2\pi}\,e^{-\frac{i\zeta}{2\pi}x_1}\frac{1}{\prod_{\alpha=1}^{N_\text{f}}2\cosh\frac{x_1-y_\alpha}{2}}.
\label{ZN1integrand}
\end{align}
Associated with each factor $\frac{1}{2\cosh\frac{x_1-y_\alpha}{2}}$, there are infinitely many poles located at
\begin{align}
x_1=y_\alpha+\pi i+2\pi in_1,\quad (n_1\in\mathbb{Z}).
\label{N1poles}
\end{align}
However, since the integrand of \eqref{ZN1integrand} is quasi-periodic under $x_1\rightarrow x_1+2\pi i$ as
\begin{align}
(\text{integrand})\Bigr|_{x_1+2\pi i} =(-1)^{N_\text{f}}e^{\zeta} (\text{integrand}),
\end{align}
we can modify the integration contour in \eqref{ZN1integrand} by introducing the extra factor from the quasi-periodicity as
\begin{align}
Z_{y_\alpha}^{(N_\text{f})}(1)=\frac{1}{1-(-1)^{N_\text{f}}e^\zeta}\frac{1}{2\cosh\frac{m}{2}}\int_\gamma\frac{dx_1}{2\pi}\,e^{-\frac{i\zeta}{2\pi}x_1}\frac{1}{\prod_{\alpha=1}^{N_\text{f}}2\cosh\frac{x_1-y_\alpha}{2}},
\label{ZN1withcountourgamma}
\end{align}
where $\gamma$ is a closed contour $(-\infty,\infty)\cup \{\infty+2\pi is\,|\,0\le s\le 1\}\cup \{-s+2\pi i\,|\,-\infty\le s\le \infty\}\cup \{-\infty+2\pi i(1-s)\,|\,0\le s\le 1\}$, as depicted in figure \ref{boxcontour}.
In this new expression of the integration, only the poles \eqref{N1poles} with $n_1=0$ are located inside the contour $\gamma$ and hence contirbute to the residue sum.
As a result, we obtain
\begin{align}
Z_{y_\alpha}^{(N_\text{f})}(1)&=\frac{1}{1-(-1)^{N_\text{f}}e^\zeta}\frac{1}{2\cosh\frac{m}{2}}\sum_{\alpha=1}^{N_\text{f}}\biggl[e^{-\frac{i\zeta}{2\pi}x_1}\frac{1}{\prod_{\beta(\neq \alpha)}^{N_\text{f}}2\cosh\frac{x_1-y_\alpha}{2}}\biggr]_{x_1=y_\alpha+\pi i}\nonumber \\
&=\frac{e^{\frac{\zeta}{2}}}{1-(-1)^{N_\text{f}}e^\zeta}\frac{1}{2\cosh\frac{m}{2}}\sum_{\alpha=1}^{N_\text{f}}R_{y_\alpha}^{-1}\frac{1}{\prod_{\beta(\neq \alpha)}^{N_\text{f}}2i\sinh\frac{y_\alpha-y_\alpha}{2}},
\end{align}
where $R_{y_\alpha}$ is defined as \eqref{RmRy}.

\subsection{$N=2$}

Next let us consider the case with $N=2$
\begin{align}
Z^{(N_\text{f})}_{y_\alpha}(2)=\frac{1}{\left(2\cosh\frac{m}{2}\right)^2}\frac{1}{2}\int_{-\infty}^\infty\frac{dx_1dx_2}{(2\pi)^2}\,e^{-\frac{i\zeta}{2\pi}(x_1+x_2)}\frac{\left(2\sinh\frac{x_1-x_2}{2}\right)^2}{\prod_\pm 2\cosh\frac{x_1-x_2\pm m}{2}\prod_{i,\alpha}2\cosh\frac{x_i-y_\alpha}{2}}.
\end{align}
In this case, the possible choices of the pole factors which determine a single point on the $x_i$-planes and the corresponding location of the poles are
\begin{subequations}
\begin{align}
&\frac{1}{2\cosh\frac{x_1-y_\alpha}{2}} \frac{1}{2\cosh\frac{x_1-x_2\pm m}{2}}\rightarrow x_1=y_\alpha+\pi i+2\pi in_1,\quad x_2=x_1\pm m+\pi i+2\pi in_2,\\
&\frac{1}{2\cosh\frac{x_2-y_\alpha}{2}} \frac{1}{2\cosh\frac{x_1-x_2\pm m}{2}}\rightarrow x_2=y_\alpha+\pi i+2\pi in_2,\quad x_1=x_2\mp m+\pi i+2\pi in_1,\\
&\frac{1}{2\cosh\frac{x_1-y_\alpha}{2}} \frac{1}{2\cosh\frac{x_2-y_\beta}{2}}\rightarrow x_1=y_\alpha+\pi i+2\pi in_1,\quad x_2=y_\beta+\pi i+2\pi in_2.\label{N2precchoice3}
\end{align}
\end{subequations}
Note that the choice of the poles with $x_1$ and $x_2$ exchanged give the identical contribution.
Hence it is sufficient to consider only one of them with the multiplicity factor precisely cancels with the overall factor $\frac{1}{2}$ of the matrix model.
Also note that for the third choice \eqref{N2precchoice3} with $\alpha=\beta$, the residue vanishes due to the factor $2\sinh\frac{x_1-x_2}{2}$ in the numerator.
Hence it is sufficient to consider only the following choices of the poles
\begin{subequations}
\begin{align}
&\text{(i)}\quad \frac{1}{2\cosh\frac{x_1-y_\alpha}{2}} \frac{1}{2\cosh\frac{x_1-x_2\pm m}{2}}\rightarrow x_1=y_\alpha+\pi i+2\pi in_1,\quad x_2=x_1\pm m+\pi i+2\pi in_2,\label{N2poles_i} \\
&\text{(iii)}\quad \frac{1}{2\cosh\frac{x_1-y_\alpha}{2}} \frac{1}{2\cosh\frac{x_2-y_\beta}{2}}\rightarrow x_1=y_\alpha+\pi i+2\pi in_1,\quad x_2=y_\beta+\pi i+2\pi in_2,\quad (\alpha<\beta).\label{N2poles_ii}
\end{align}
\end{subequations}
As was the case for $N=1$, by taking into account of the quasi-periodicity of the integrand we can simplify the residue sum over infinitely many poles to the summation over the poles with $n_i=0$ for each choice (i) and (ii)
\begin{align}
Z^{(N_\text{f})}_{y_\alpha}(2)=\sum_\alpha \sum_\pm Z^{(N_\text{f})}_{y_\alpha}(2;\text{(i)}_{\alpha,\pm})+\sum_{\alpha<\beta} Z^{(N_\text{f})}_{y_\alpha}(2;\text{(ii)}_{\alpha,\beta}),
\end{align}
where $Z^{(N_\text{f})}_{y_\alpha}(2;\text{(i)}_{\alpha,\pm})$ and $Z^{(N_\text{f})}_{y_\alpha}(2;\text{(ii)}_{\alpha,\beta})$ are respectively the contribution from the choice (i) with a fixed $(\alpha,\pm)$ and the contribution from the choice (ii) with a fixed $(\alpha,\beta)$.
Note, however, that the quasi-periodicity factor can change as we perform the integration over part of the $N$ integration variables.
For concreteness, in both (i) and (ii) let us perform the integration over $x_2$ first and then perform the integration over $x_1$.
In the choice (i), when we integrate over $x_2$ the quasi-periodicity of the integrand is
\begin{align}
(\text{integrand})\Bigr|_{x_2+2\pi i} =(-1)^{N_\text{f}}e^{\zeta} (\text{integrand}).
\end{align}
On the other hand, after performing the integration, $x_2$ in the FI term and the one-loop determinant of the fundamental hypermultiplets are relpaced with $x_1+(const.)$.
Hence when we integrate over $x_1$, the quasi-periodicity of the integrand is modified as
\begin{align}
(\text{integrand})\Bigr|_{x_1+2\pi i} =e^{2\zeta} (\text{integrand}).
\end{align}
As a result, the contribution from the choice (i) with a fixed $\alpha$ and $\pm$ to the partition function is
\begin{align}
&Z^{(N_\text{f})}_{y_\alpha}(2;\text{(i)}_{\alpha,\pm}) =
\frac{1}{1-e^{2\zeta}}\frac{1}{1-(-1)^{N_\text{f}}e^\zeta}
\frac{1}{\left(2\cosh\frac{m}{2}\right)^2}\nonumber \\
&\quad\times \biggl[
e^{-\frac{i\zeta}{2\pi}(x_1+x_2)}\frac{\left(2\sinh\frac{x_1-x_2}{2}\right)^2}{2\cosh\frac{x_1-x_2\mp m}{2}
\prod_{\beta(\neq\alpha)}2\cosh\frac{x_1-y_\beta}{2}
\prod_{\beta}2\cosh\frac{x_2-y_\beta}{2}
}
\biggr]_{\substack{x_1=y_\alpha+\pi i,\\
x_2=x_1\pm m+\pi i
}
}\nonumber \\
&=
\frac{\mp i^{-N_\text{f}}}{(2\sinh m)(2\cosh\frac{m}{2})(2\sinh\zeta)\left(e^{\frac{\zeta}{2}}-(-1)^{N_\text{f}}e^{-\frac{\zeta}{2}}\right)}
\frac{R_m^{\mp 1}R_{y_\alpha}^{-2}}{
\prod_{\beta(\neq\alpha)}2\sinh\frac{y_\alpha-y_\beta}{2}2\cosh\frac{y_\alpha-y_\beta\pm m}{2}
},\label{ZNf;1(2)}
\end{align}
where $R_m$ is defined as \eqref{RmRy}.

In the choice (ii), the integration over $x_2$ does not affect the quasi-periodicity with respect to $x_1$
\begin{subequations}
\begin{align}
&(\text{integrand})\Bigr|_{x_2+2\pi i} =(-1)^{N_\text{f}}e^{\zeta} (\text{integrand}),\\
&(\text{integrand})\Bigr|_{x_1+2\pi i} =(-1)^{N_\text{f}}e^{\zeta} (\text{integrand}).
\end{align}
\end{subequations}
Hence we obtain
\begin{align}
&Z^{(N_\text{f})}_{y_\alpha}(2;\text{(ii)}_{\alpha,\beta})
=\left(\frac{1}{1-(-1)^{N_\text{f}}e^\zeta}\right)^2\frac{1}{\left(2\cosh\frac{m}{2}\right)^2}\nonumber \\
&\quad\times \biggl[
e^{-\frac{i\zeta}{2\pi}(x_1+x_2)}\frac{\left(2\sinh\frac{x_1-x_2}{2}\right)^2}{\prod_\pm 2\cosh\frac{x_1-x_2\pm m}{2}
\prod_{\gamma(\neq\alpha)}2\cosh\frac{x_1-y_\gamma}{2}
\prod_{\delta(\neq\beta)}2\cosh\frac{x_2-y_\delta}{2}
}
\biggr]_{\substack{
x_1=y_\alpha+\pi i,\\
x_2=y_\beta+\pi i
}}\nonumber \\
&=\frac{(-1)^{N_\text{f}}}{(2\cosh\frac{m}{2})^2\left(e^{\frac{\zeta}{2}}-(-1)^{N_\text{f}}e^{-\frac{\zeta}{2}}\right)^2}\frac{R_{y_\alpha}R_{y_\beta}}{\prod_\pm 2\cosh\frac{y_\alpha-y_\beta\pm m}{2}\prod_{\gamma(\neq \alpha,\beta)}2\sinh\frac{y_\alpha-y_\gamma}{2}2\sinh\frac{y_\beta-y_\gamma}{2}}.
\label{ADHMgeneralNfexactvaluesN2}
\end{align}

\subsection{$N=3$}

Lastly, let us consider the case with $N=3$
\begin{align}
&Z^{(N_\text{f})}_{y_\alpha}(3)=\frac{1}{\left(2\cosh\frac{m}{2}\right)^3}\frac{1}{6}\int_{-\infty}^\infty\frac{dx_1dx_2dx_3}{(2\pi)^3}\, e^{-\frac{i\zeta}{2\pi}(x_1+x_2+x_3)}\nonumber \\
&\quad\times \frac{\left(2\sinh\frac{x_1-x_2}{2}\right)^2\left(2\sinh\frac{x_1-x_3}{2}\right)^2\left(2\sinh\frac{x_2-x_3}{2}\right)^2}{\prod_\pm
2\cosh\frac{x_1-x_2\pm m}{2}
2\cosh\frac{x_1-x_3\pm m}{2}
2\cosh\frac{x_2-x_3\pm m}{2}
\prod_{i,\alpha}2\cosh\frac{x_i-y_\alpha}{2}}.
\end{align}
In this case, the possible choices of the pole factors up to the permutation of $\{x_1,x_2,x_3\}$ are
\begin{subequations}
\begin{align}
&\frac{1}{2\cosh\frac{x_1-y_\alpha}{2}}\frac{1}{2\cosh\frac{x_1-x_2\pm m}{2}}\frac{1}{2\cosh\frac{x_2-x_3\pm' m}{2}}\nonumber \\
&\quad \rightarrow x_1=y_\alpha+\pi i+2\pi in_1,\quad
x_2=x_1\pm m+\pi i+2\pi in_2,\quad
x_3=x_2\pm' m+\pi i+2\pi in_3,\\
&\frac{1}{2\cosh\frac{x_1-y_\alpha}{2}}\frac{1}{2\cosh\frac{x_1-x_2\pm m}{2}}\frac{1}{2\cosh\frac{x_1-x_3\pm' m}{2}}\quad ((\pm,\pm')=(++),(+-),(--))\nonumber \\
&\quad \rightarrow x_1=y_\alpha+\pi i+2\pi in_1,\quad
x_2=x_1\pm m+\pi i+2\pi in_2,\quad
x_3=x_1\pm' m+\pi i+2\pi in_3,\\
&\frac{1}{2\cosh\frac{x_1-y_\alpha}{2}}\frac{1}{2\cosh\frac{x_2-y_\beta}{2}}\frac{1}{2\cosh\frac{x_1-x_3\pm m}{2}}\nonumber \\
&\quad \rightarrow x_1=y_\alpha+\pi i+2\pi in_1,\quad
x_2=y_\beta+\pi i+2\pi in_2,\quad
x_3=x_1\pm m+\pi i+2\pi in_3,\\
&\frac{1}{2\cosh\frac{x_1-y_\alpha}{2}}\frac{1}{2\cosh\frac{x_2-y_\beta}{2}}\frac{1}{2\cosh\frac{x_3-y_\gamma}{2}}\quad (\alpha\le \beta\le \gamma)\nonumber \\
&\quad \rightarrow x_1=y_\alpha+\pi i+2\pi in_1,\quad
x_2=y_\beta+\pi i+2\pi in_2,\quad
x_3=y_\gamma+\pi i+2\pi in_3.
\end{align}
\end{subequations}
Again, the second choice with $(\pm,\pm')=(++),(--)$ and the fourth choice with $\alpha=\beta$ or $\beta=\gamma$ have vanishing residues and hence we do not have to consider these choices.
Furthermore, the first choice with $(\pm,\pm')=(+-),(-+)$ also have vanishing residues.
Indeed, substituting $x_2=x_1\pm m+\pi i+2\pi in_2$ to $x_3$ we have $x_3=x_1+(\pm 1 \pm' 1)m+2\pi i(1+n_2+n_3)$.
When $(\pm,\pm')=(+-)$ or $(-+)$ this makes the factor $2\sinh\frac{x_1-x_3}{2}$ in the numerator vanish.
After all, the choices of the poles we have to take into account are only the following four
\begin{subequations}
\begin{align}
&\text{(i)}\quad \frac{1}{2\cosh\frac{x_1-y_\alpha}{2}}\frac{1}{2\cosh\frac{x_1-x_2\pm m}{2}}\frac{1}{2\cosh\frac{x_2-x_3\pm m}{2}}\nonumber \\
&\quad\quad \rightarrow x_1=y_\alpha+\pi i+2\pi in_1,\quad
x_2=x_1\pm m+\pi i+2\pi in_2,\quad
x_3=x_2\pm m+\pi i+2\pi in_3,\label{N3poles_i}\\
&\text{(ii)}\quad \frac{1}{2\cosh\frac{x_1-y_\alpha}{2}}\frac{1}{2\cosh\frac{x_1-x_2+m}{2}}\frac{1}{2\cosh\frac{x_1-x_3-m}{2}}\nonumber \\
&\quad\quad \rightarrow x_1=y_\alpha+\pi i+2\pi in_1,\quad
x_2=x_1+m+\pi i+2\pi in_2,\quad
x_3=x_1-m+\pi i+2\pi in_3,\label{N3poles_ii}\\
&\text{(iii)}\quad \frac{1}{2\cosh\frac{x_1-y_\alpha}{2}}\frac{1}{2\cosh\frac{x_2-y_\beta}{2}}\frac{1}{2\cosh\frac{x_1-x_3\pm m}{2}}\quad (\alpha\neq \beta)\nonumber \\
&\quad\quad \rightarrow x_1=y_\alpha+\pi i+2\pi in_1,\quad
x_2=y_\beta+\pi i+2\pi in_2,\quad
x_3=x_1\pm m+\pi i+2\pi in_3,\label{N3poles_iii}\\
&\text{(iv)}\quad \frac{1}{2\cosh\frac{x_1-y_\alpha}{2}}\frac{1}{2\cosh\frac{x_2-y_\beta}{2}}\frac{1}{2\cosh\frac{x_3-y_\gamma}{2}}\quad (\alpha<\beta<\gamma)\nonumber \\
&\quad\quad \rightarrow x_1=y_\alpha+\pi i+2\pi in_1,\quad
x_2=y_\beta+\pi i+2\pi in_2,\quad
x_3=y_\gamma+\pi i+2\pi in_3.\label{N3poles_iv}
\end{align}
\end{subequations}
Note that in (iii) we have to take into account of both $\alpha>\beta$ and $\alpha<\beta$ since they are not related to each other by the permutation of $\{x_1,x_2,x_3\}$.
Hence we have
\begin{align}
Z^{(N_\text{f})}_{y_\alpha}(3)&=\sum_\alpha \sum_\pm Z^{(N_\text{f})}_{y_\alpha}(3;\text{(i)}_{\alpha,\pm})
+\sum_{\alpha} Z^{(N_\text{f})}_{y_\alpha}(3;\text{(ii)}_{\alpha})
+\sum_{\alpha\neq \beta}\sum_\pm Z^{(N_\text{f})}_{y_\alpha}(3;\text{(iii)}_{\alpha,\beta,\pm})\nonumber \\
& \quad +\sum_{\alpha<\beta<\gamma} Z^{(N_\text{f})}_{y_\alpha}(3;\text{(iv)}_{\alpha,\beta,\gamma}),
\end{align}
where $Z^{(N_\text{f})}_{y_\alpha}(3;\text{(i)}_{\alpha,\pm})$, etc.~are the contribution from each choice of the poles.
The contributions from the infinite towers of the poles generated by $n_1,n_2,n_3$ are taken care of by the quasi-periodicity factors, which we can calculate explicitly by choosing the order of the integrations, say $x_3\rightarrow x_2\rightarrow x_1$.
For (i), the quasi-periodicity of the integrand changes as we perform the integrations, as
\begin{subequations}
\begin{align}
&(\text{integrand})\Bigr|_{x_3+2\pi i} = (-1)^{N_\text{f}}e^\zeta (\text{integrand}),\\
&(\text{integrand})\Bigr|_{x_2+2\pi i} = e^{2\zeta} (\text{integrand}),\\
&(\text{integrand})\Bigr|_{x_1+2\pi i} = (-1)^{N_\text{f}}e^{3\zeta} (\text{integrand}).
\end{align}
\end{subequations}
Hence the net contribution from the poles (i) with fixed $(\alpha,\pm)$ is
\begin{align}
& Z^{(N_\text{f})}_{y_\alpha}(3;\text{(i)}_{\alpha,\pm}) =
\frac{1}{1-(-1)^{N_\text{f}}e^{3\zeta}}
\frac{1}{1-e^{2\zeta}}
\frac{1}{1-(-1)^{N_\text{f}}e^{\zeta}}
\frac{1}{\left(2\cosh\frac{m}{2}\right)^3}
\Biggr[
e^{-\frac{i\zeta}{2\pi}(x_1+x_2+x_3)} \nonumber \\
&\quad \quad \times \frac{\left(2\sinh\frac{x_1-x_2}{2}\right)^2\left(2\sinh\frac{x_1-x_3}{2}\right)^2\left(2\sinh\frac{x_2-x_3}{2}\right)^2}{
2\cosh\frac{x_1-x_2\mp m}{2}
2\cosh\frac{x_2-x_3\mp m}{2}
\prod_{\pm'}
2\cosh\frac{x_1-x_3\pm' m}{2}
\prod_{\beta(\neq\alpha)}2\cosh\frac{x_1-y_\beta}{2}
}\nonumber \\
&\quad\quad \quad \quad \quad \quad \quad \quad \quad \quad \times \prod_{\beta}\frac{1}{2\cosh\frac{x_2-y_\beta}{2}2\cosh\frac{x_3-y_\beta}{2}}\Biggr]_{\substack{x_1=y_\alpha+\pi i,\\
x_2=x_1\pm m+\pi i,\\
x_3=x_2\pm m+\pi i}}\nonumber \\
&\quad =
\frac{\pm i(-1)^{N_\text{f}}}{(2\cosh\frac{3m}{2})(2\sinh m)(2\cosh\frac{m}{2})}
\frac{1}{e^{\frac{3\zeta}{2}}-(-1)^{N_\text{f}}e^{-\frac{3\zeta}{2}}}
\frac{1}{2\sinh \zeta}
\frac{1}{e^{\frac{\zeta}{2}}-(-1)^{N_\text{f}}e^{-\frac{\zeta}{2}}}\nonumber \\
&\quad\quad \quad \quad \quad \times \frac{R_m^{\mp 3}R_{y_\alpha}^{-3}}{\prod_{\beta(\neq\alpha)}2\sinh\frac{y_\alpha-y_\beta}{2}2\cosh\frac{y_\alpha-y_\beta\pm m}{2}2\sinh\frac{y_\alpha-y_\beta\pm 2m}{2}}.\label{ZNf;13(3)}
\end{align}
For (ii), the quasi-periodicity of the integrand changes as we perform the integrations, as
\begin{subequations}
\begin{align}
&(\text{integrand})\Bigr|_{x_3+2\pi i} = (-1)^{N_\text{f}}e^\zeta (\text{integrand}),\\
&(\text{integrand})\Bigr|_{x_2+2\pi i} = (-1)^{N_\text{f}}e^\zeta (\text{integrand}),\\
&(\text{integrand})\Bigr|_{x_1+2\pi i} = (-1)^{N_\text{f}}e^{3\zeta} (\text{integrand}).
\end{align}
\end{subequations}
Hence the net contribution from the poles (ii) with fixed $\alpha$ is
\begin{align}
&Z^{(N_\text{f})}_{y_\alpha}(3;\text{(ii)}_{\alpha}) =
\frac{1}{1-(-1)^{N_\text{f}}e^{3\zeta}}
\left(
\frac{1}{1-(-1)^{N_\text{f}}e^{\zeta}}
\right)^2
\frac{1}{\left(2\cosh\frac{m}{2}\right)^3}
\biggl[
e^{-\frac{i\zeta}{2\pi}(x_1+x_2+x_3)}\nonumber \\
&\quad \times \frac{\left(2\sinh\frac{x_1-x_2}{2}\right)^2\left(2\sinh\frac{x_1-x_3}{2}\right)^2\left(2\sinh\frac{x_2-x_3}{2}\right)^2}{
2\cosh\frac{x_1-x_2- m}{2}
2\cosh\frac{x_1-x_3+ m}{2}
\prod_{\pm}
2\cosh\frac{x_2-x_3\pm m}{2}
\prod_{\beta(\neq\alpha)}2\cosh\frac{x_1-y_\beta}{2}
}\nonumber \\
&\quad \quad \quad \quad \quad \quad \times \prod_{\beta}\frac{1}{2\cosh\frac{x_2-y_\beta}{2}2\cosh\frac{x_3-y_\beta}{2}}\Biggr]_{\substack{x_1=y_\alpha+\pi i,\\
x_2=x_1+ m+\pi i,\\
x_3=x_1- m+\pi i}}\nonumber \\
&\quad =
\frac{1}{(2\cosh\frac{3m}{2})(2\cosh\frac{m}{2})^2}\frac{i^{N_\text{f}-1}}{e^{\frac{3\zeta}{2}}-(-1)^{N_\text{f}}e^{-\frac{3\zeta}{2}}}
\frac{1}{\left(e^{\frac{\zeta}{2}}-(-1)^{N_\text{f}}e^{-\frac{\zeta}{2}}\right)^2}\nonumber \\
&\quad \quad \quad \quad \quad\times \frac{R_{y_\alpha}^{-3}}{\prod_{\beta(\neq \alpha)}2\sinh\frac{y_\alpha-y_\beta}{2}\prod_\pm 2\cosh\frac{y_\alpha-y_\beta\pm m}{2}}.
\end{align}
For (iii), the quasi-periodicity of the integrand changes as we perform the integrations, as
\begin{subequations}
\begin{align}
&(\text{integrand})\Bigr|_{x_3+2\pi i} = (-1)^{N_\text{f}}e^\zeta (\text{integrand}),\\
&(\text{integrand})\Bigr|_{x_2+2\pi i} = (-1)^{N_\text{f}}e^\zeta (\text{integrand}),\\
&(\text{integrand})\Bigr|_{x_1+2\pi i} = e^{2\zeta} (\text{integrand}).
\end{align}
\end{subequations}
Hence the net contribution from the poles (iii) with fixed $(\alpha,\beta,\pm)$ is
\begin{align}
&Z^{(N_\text{f})}_{y_\alpha}(3;\text{(iii)}_{\alpha,\beta,\pm}) =
\frac{1}{1-e^{2\zeta}}
\left(
\frac{1}{1-(-1)^{N_\text{f}}e^{\zeta}}
\right)^2
\frac{1}{\left(2\cosh\frac{m}{2}\right)^3}
\biggl[
e^{-\frac{i\zeta}{2\pi}(x_1+x_2+x_3)}\nonumber \\
&\quad \times \frac{\left(2\sinh\frac{x_1-x_2}{2}\right)^2\left(2\sinh\frac{x_1-x_3}{2}\right)^2\left(2\sinh\frac{x_2-x_3}{2}\right)^2}{
2\cosh\frac{x_1-x_3\mp m}{2}
\prod_{\pm'}
2\cosh\frac{x_1-x_2\pm' m}{2}
2\cosh\frac{x_2-x_3\pm' m}{2}
\prod_{\gamma(\neq\alpha)}2\cosh\frac{x_1-y_\gamma}{2}
}\nonumber \\
&\quad\quad \quad \quad \quad \times 
\prod_{\delta(\neq\beta)}\frac{1}{2\cosh\frac{x_2-y_\delta}{2}}
\prod_{\gamma}\frac{1}{2\cosh\frac{x_3-y_\gamma}{2}}\biggr]_{\substack{x_1=y_\alpha+\pi i,\\
x_2=y_\beta +\pi i,\\
x_3=x_1\pm m+\pi i}}\nonumber \\
&\quad \quad \quad \quad =
\frac{\mp i}{(2\sinh m)(2\cosh\frac{m}{2})^2(2\sinh\zeta)}\frac{1}{\left(e^{\frac{\zeta}{2}}-(-1)^{N_\text{f}}e^{-\frac{\zeta}{2}}\right)^2}\nonumber \\
&\quad\quad \quad \quad \quad \times \frac{R_{y_\alpha}^{-2}R_{y_\beta}^{-1}R_m^{\mp 1}}{\prod_{\gamma(\neq \alpha,\beta)}2\sinh\frac{y_\alpha-y_\gamma}{2}2\sinh\frac{y_\beta-y_\gamma}{2}2\cosh\frac{y_\alpha-y_\gamma\pm m}{2}}\nonumber \\
& \quad \quad \quad \quad \quad \quad \times \frac{1}{2\sinh\frac{y_\alpha-y_\beta}{2}2\cosh\frac{y_\alpha-y_\beta\mp m}{2}2\sinh\frac{y_\alpha-y_\beta\pm 2m}{2}}.
\end{align}
For (iv), the quasi-periodicity of the integrand does not change as we perform the integrations
\begin{subequations}
\begin{align}
&(\text{integrand})\Bigr|_{x_3+2\pi i} = (-1)^{N_\text{f}}e^\zeta (\text{integrand}),\\
&(\text{integrand})\Bigr|_{x_2+2\pi i} = (-1)^{N_\text{f}}e^\zeta (\text{integrand}),\\
&(\text{integrand})\Bigr|_{x_1+2\pi i} = (-1)^{N_\text{f}}e^\zeta (\text{integrand}).
\end{align}
\end{subequations}
Hence the net contribution from the poles (iv) with fixed $(\alpha,\beta,\gamma)$ is
\begin{align}
&Z^{(N_\text{f})}_{y_\alpha}(3;\text{(iv)}_{\alpha,\beta,\gamma}) =
\left(
\frac{1}{1-(-1)^{N_\text{f}}e^{\zeta}}
\right)^3
\frac{1}{\left(2\cosh\frac{m}{2}\right)^3}
\biggl[
e^{-\frac{i\zeta}{2\pi}(x_1+x_2+x_3)}\nonumber \\
&\quad\quad\times \frac{\left(2\sinh\frac{x_1-x_2}{2}\right)^2\left(2\sinh\frac{x_1-x_3}{2}\right)^2\left(2\sinh\frac{x_2-x_3}{2}\right)^2}{
\prod_{\pm}
2\cosh\frac{x_1-x_2\pm m}{2}
2\cosh\frac{x_1-x_3\pm m}{2}
2\cosh\frac{x_2-x_3\pm m}{2}
\prod_{\delta(\neq\alpha)}2\cosh\frac{x_1-y_\delta}{2}
}\nonumber \\
&\quad\quad\quad\quad\quad\quad\times 
\prod_{\delta'(\neq\beta)}\frac{1}{2\cosh\frac{x_2-y_{\delta'}}{2}}
\prod_{\delta''(\neq\gamma)}\frac{1}{2\cosh\frac{x_3-y_{\delta''}}{2}}
\biggr]_{\substack{x_1=y_\alpha+\pi i,\\
x_2=y_\beta +\pi i,\\
x_3=y_\gamma+\pi i}}\nonumber \\
&=
\frac{1}{\left(2\cosh\frac{m}{2}\right)^3}\frac{i^{3N_\text{f}-1}}{\left(e^{\frac{\zeta}{2}}-(-1)^{N_\text{f}}e^{-\frac{\zeta}{2}}\right)^3}
\frac{
R_{y_\alpha}^{-1}
R_{y_\beta}^{-1}
R_{y_\gamma}^{-1}
}{\prod_\pm 2\cosh\frac{y_\alpha-y_\beta\pm m}{2}
2\cosh\frac{y_\alpha-y_\gamma\pm m}{2}
2\cosh\frac{y_\beta-y_\gamma\pm m}{2}
}\nonumber \\
&\quad\quad\quad\quad\quad\quad\times \frac{1}{\prod_{\delta(\neq \alpha,\beta,\gamma)}
2\sinh\frac{y_\alpha-y_\delta}{2}
2\sinh\frac{y_\beta-y_\delta}{2}
2\sinh\frac{y_\gamma-y_\delta}{2}
}.
\label{ZNf;6}
\end{align}

Note that the choice of the poles in the above calculations for $N=1,2,3$ are labelled by $N_\text{f}$-ples of Young diagrams $(\lambda^{(1)},\lambda^{(2)},\cdots,\lambda^{(N_\text{f})})$ satisfying $\sum_{\alpha=1}^{N_\text{f}}|\lambda^{(\alpha)}|=N$.
Hence total partition function $Z_{y_\alpha}^{(N_\text{f})}(N)$ is given by a summation over the Young diagrams
\begin{align}
Z_{y_\alpha}^{(N_\text{f})}(N)=\!\!\!\!\!\!\!\!\!\!\! \sum_{\substack{\lambda^{(\alpha)}\\ \left(\sum_{\alpha=1}^{N_\text{f}}|\lambda^{(\alpha)}|=N\right)}} \!\!\!\!\!\!\!\!\!\!\! Z_{\lambda^{(\alpha)}}^{(N_\text{f})}.
\label{totalZincombinatorialformula}
\end{align}
See appendix \ref{app_Youngdiagramnotation} for various notation for a Young diagram.
Here the Young diagrams are drawn in the following rule
\begin{itemize}
\item For each factor $\frac{1}{2\cosh\frac{x_i-y_\alpha}{2}}$, draw a top-left box of the $\alpha$-th Young diagram and assign the variable $x_i$ to that box.
\item For each factor $\frac{1}{2\cosh\frac{x_i-x_j+m}{2}}$, draw a box on the right of the box assigned with $x_i$, and assign $x_j$ to the new box.
\item For each factor $\frac{1}{2\cosh\frac{x_i-x_j-m}{2}}$, draw a box below the box assigned with $x_i$, and assign $x_j$ to the new box.
\end{itemize}
In table \ref{listofYforNf3N123} we list the explicit relation between the pole factors and the Young diagrams for $N_\text{f}=3$ and $N\le 3$.
\begin{table}
\begin{center}
\begin{tabular}{|c|c|c|}
\hline
$N$&$(\lambda^{(1)},\lambda^{(2)},\lambda^{(3)})$&$\text{pole}$\\ \hline
$1$&$(\ydiagram{1},\emptyset,\emptyset)$                         &\eqref{N1poles} with $\alpha=1$\\ \cline{2-3}
   &$(\emptyset,\ydiagram{1},\emptyset)$                         &\eqref{N1poles} with $\alpha=2$\\ \cline{2-3}
   &$(\emptyset,\emptyset,\ydiagram{1})$                         &\eqref{N1poles} with $\alpha=3$\\ \hline
$2$&$(\ydiagram{2},\emptyset,\emptyset)$                         &(i) \eqref{N2poles_i} with $(\alpha,\pm)=(1,+)$\\ \cline{2-3}
   &$(\ydiagram{1,1},\emptyset ,\emptyset)$                      &(i) \eqref{N2poles_i} with $(\alpha,\pm)=(1,-)$\\ \cline{2-3}
   &$(\emptyset,\ydiagram{2},\emptyset)$                         &(i) \eqref{N2poles_i} with $(\alpha,\pm)=(2,+)$\\ \cline{2-3}
   &$(\emptyset,\ydiagram{1,1},\emptyset)$                       &(i) \eqref{N2poles_i} with $(\alpha,\pm)=(2,-)$\\ \cline{2-3}
   &$(\emptyset,\emptyset,\ydiagram{2})$                         &(i) \eqref{N2poles_i} with $(\alpha,\pm)=(3,+)$\\ \cline{2-3}
   &$(\emptyset,\emptyset,\ydiagram{1,1})$                       &(i) \eqref{N2poles_i} with $(\alpha,\pm)=(3,-)$\\ \cline{2-3}
   &$(\ydiagram{1},\ydiagram{1},\emptyset)$                      &(ii) \eqref{N2poles_ii} with $(\alpha,\beta)=(1,2)$\\ \cline{2-3}
   &$(\ydiagram{1},\emptyset,\ydiagram{1})$                      &(ii) \eqref{N2poles_ii} with $(\alpha,\beta)=(1,3)$\\ \cline{2-3}
   &$(\emptyset,\ydiagram{1},\ydiagram{1})$                      &(ii) \eqref{N2poles_ii} with $(\alpha,\beta)=(2,3)$\\ \hline
$3$&$(\ydiagram{3},\emptyset     ,\emptyset)$                              &(i) \eqref{N3poles_i} with $(\alpha,\pm)=(1,+)$\\ \cline{2-3}
   &$(\ydiagram{2,1},\emptyset   ,\emptyset)$                              &(ii) \eqref{N3poles_ii} with $\alpha=1$\\ \cline{2-3}
   &$(\ydiagram{1,1,1},\emptyset ,\emptyset)$                              &(i) \eqref{N3poles_i} with $(\alpha,\pm)=(1,-)$\\ \cline{2-3}
   &$(\emptyset,\ydiagram{3}     ,\emptyset)$                              &(i) \eqref{N3poles_i} with $(\alpha,\pm)=(2,+)$\\ \cline{2-3}
   &$(\emptyset,\ydiagram{2,1}   ,\emptyset)$                              &(ii) \eqref{N3poles_ii} with $\alpha=2$\\ \cline{2-3}
   &$(\emptyset,\ydiagram{1,1,1} ,\emptyset)$                              &(i) \eqref{N3poles_i} with $(\alpha,\pm)=(2,-)$\\ \cline{2-3}
   &$(\emptyset,\emptyset,\ydiagram{3}    )$                               &(i) \eqref{N3poles_i} with $(\alpha,\pm)=(3,+)$\\ \cline{2-3}
   &$(\emptyset,\emptyset,\ydiagram{2,1}  )$                               &(ii) \eqref{N3poles_ii} with $\alpha=3$\\ \cline{2-3}
   &$(\emptyset,\emptyset,\ydiagram{1,1,1})$                               &(i) \eqref{N3poles_i} with $(\alpha,\pm)=(3,-)$\\ \cline{2-3}
   &$(\ydiagram{2},\ydiagram{1}  ,\emptyset)$                              &(iii) \eqref{N3poles_iii} with $(\alpha,\beta,\pm)=(1,2,+)$\\ \cline{2-3}
   &$(\ydiagram{1,1},\ydiagram{1},\emptyset)$                              &(iii) \eqref{N3poles_iii} with $(\alpha,\beta,\pm)=(1,2,-)$\\ \cline{2-3}
   &$(\ydiagram{1},\ydiagram{2}  ,\emptyset)$                              &(iii) \eqref{N3poles_iii} with $(\alpha,\beta,\pm)=(2,1,+)$\\ \cline{2-3}
   &$(\ydiagram{1},\ydiagram{1,1},\emptyset)$                              &(iii) \eqref{N3poles_iii} with $(\alpha,\beta,\pm)=(2,1,-)$\\ \cline{2-3}
   &$(\ydiagram{2},  \emptyset,\ydiagram{1})$                              &(iii) \eqref{N3poles_iii} with $(\alpha,\beta,\pm)=(1,3,+)$\\ \cline{2-3}
   &$(\ydiagram{1,1},\emptyset,\ydiagram{1})$                              &(iii) \eqref{N3poles_iii} with $(\alpha,\beta,\pm)=(1,3,-)$\\ \cline{2-3}
   &$(\ydiagram{1},  \emptyset,\ydiagram{2})$                              &(iii) \eqref{N3poles_iii} with $(\alpha,\beta,\pm)=(3,1,+)$\\ \cline{2-3}
   &$(\ydiagram{1},  \emptyset,\ydiagram{1,1})$                            &(iii) \eqref{N3poles_iii} with $(\alpha,\beta,\pm)=(3,1,-)$\\ \cline{2-3}
   &$(\emptyset,\ydiagram{2},\ydiagram{1}  )$                              &(iii) \eqref{N3poles_iii} with $(\alpha,\beta,\pm)=(2,3,+)$\\ \cline{2-3}
   &$(\emptyset,\ydiagram{1,1},\ydiagram{1})$                              &(iii) \eqref{N3poles_iii} with $(\alpha,\beta,\pm)=(2,3,-)$\\ \cline{2-3}
   &$(\emptyset,\ydiagram{1},\ydiagram{2}  )$                              &(iii) \eqref{N3poles_iii} with $(\alpha,\beta,\pm)=(3,2,+)$\\ \cline{2-3}
   &$(\emptyset,\ydiagram{1},\ydiagram{1,1})$                              &(iii) \eqref{N3poles_iii} with $(\alpha,\beta,\pm)=(3,2,-)$\\ \cline{2-3}
   &$(\ydiagram{1},\ydiagram{1},\ydiagram{1})$                             &(iv) \eqref{N3poles_iv} with $(\alpha,\beta,\gamma)=(1,2,3)$\\ \hline
\end{tabular}
\caption{
List of Young diagrams $(\lambda^{(1)},\lambda^{(2)},\lambda^{(3)})$ associated with each choice of the poles for $N_\text{f}=3$ and $N=1,2,3$.
}
\label{listofYforNf3N123}
\end{center}
\end{table}
In particular, the powers of $R_m$ and $R_{y_\alpha}$ for each $Z^{(N_\text{f})}_{\lambda^{(\alpha)}}$ can be estimated in a simple way as
\begin{subequations}
\label{powerofRmandRy}
\begin{align}
Z^{(N_\text{f})}_{\lambda^{(\alpha)}}&\propto \prod_{\alpha=1}^{N_\text{f}}R_m^{-\sum_{\Box=(a,b)\in\lambda^{(\alpha)}}(b-a)}=R_m^{-\frac{1}{2}\sum_{\alpha=1}^{N_\text{f}}\sum_{i=1}^{\ell(\lambda^{(\alpha)})}\lambda_i^{(\alpha)}\left(\lambda_i^{(\alpha)}-2i+1\right)},\\
Z^{(N_\text{f})}_{\lambda^{(\alpha)}}&\propto \prod_{\alpha=1}^{N_\text{f}}R_{y_\alpha}^{-|\lambda^{(\alpha)}|}.
\end{align}
\end{subequations}
Here the first formula follows from the fact that for each connected Young diagram $\lambda^{(\alpha)}$, after performing the residue calculation the variable $x_i$ associated with the box $\Box=(a,b)\in\lambda$ is replaced with $x_{\text{top-left}}+(b-a)m$ up to the shift by $\pi i$ times some integer. 
Similarly, the second formula follows from the fact that after this replacement we are left with $e^{-\frac{i\zeta}{2\pi}(|\lambda^{(\alpha)}|x_{\text{top-lef}}+\cdots)}$ in the exponent of the FI term, and then $x_{\text{top-left}}$ is substituted with $y_\alpha$.

Conversely, if we know the full partition function $Z_{y_\alpha}^{(N_\text{f})}(N)$ via some different method such as the Fermi gas formalism \cite{Nosaka:2020tyv}, we can read off the building blocks $Z_{\lambda^{(\alpha)}}^{(N_\text{f})}$ unless there is a degeneracy in the correspondence between the powers of $(R_m,R_{y_\alpha})$ and the Young diagrams $(\lambda^{(1)},\cdots,\lambda^{(N_\text{f})})$.
We will use this point in section \ref{sec_evidenceforcombinatorialformula}.

\section{Combinatorial formulas for general $N$}
\label{sec_generalcombinatorial}

The partition function with rank $N\ge 4$ can also be calculated by the residue sum where the choice of the poles are associated with the Young diagram through the rule described above.
When all of the Young diagrams are of hook type, the above calculation of the building blocks $Z_{\lambda^{(\alpha)}}^{(N_\text{f})}$ can be generalized straightforwardly.
In particular, by performing the integration from the variables associated with right/bottom boxes to top/left boxes, we find that the quasi-periodicity of the integrand when we perform the integration over $x_i$ which is associated with a box $\Box\in\lambda^{(\alpha)}$ is
\begin{align}
(\text{integrand})\Bigr|_{x_i+2\pi i}&=((-1)^{N_\text{f}}e^\zeta)^{\#(\text{box on the right/bottom of }\Box)}(\text{integrand})\nonumber \\
&=((-1)^{N_\text{f}}e^\zeta)^{h_{\lambda^{(\alpha)}}(\Box)}(\text{integrand}),
\label{quasiperiodicityfactorgeneral}
\end{align}
where $h_\lambda(\Box)$ is the hook length \eqref{hook} at $\Box\in\lambda$.
Hence we conclude
\begin{align}
&Z^{(N_\text{f})}_{\lambda^{(\alpha)}}=
\Biggl(
\prod_{\alpha=1}^{N_\text{f}}\prod_{\Box\in\lambda^{(\alpha)}}\frac{1}{\displaystyle 1-((-1)^{N_\text{f}}e^{\zeta})^{h_{\lambda^{(\alpha)}}(\Box)}}\Biggr)
\frac{1}{(2\cosh\frac{m}{2})^N}
\nonumber \\
&\quad\times \left[
e^{\frac{\zeta}{2\pi i}\sum_i x_i}\frac{\displaystyle {\prod}_{i<j}\Bigl(2\sinh\frac{x_i-x_j}{2}\Bigr)^2}{\displaystyle {\prod}^\prime_{\substack{i<j,\pm}} 2\cosh\frac{x_i-x_j\pm m}{2}}
\frac{1}{\displaystyle {\prod}^\prime_{\substack{i,\alpha}} 2\cosh\frac{x_i-y_\alpha}{2}}
\right]_{\text{pole}},
\label{ZNflambdaalpha2kaime}
\end{align}
where the notation ${\prod}^\prime$ means the product without pole factors.
A more careful analysis also allows us to write down the closed form expression of the whole building block $Z_{\lambda^{(\alpha)}}^{(N_\text{f})}$.
When $N_\text{f}=1$, we have \cite{Gaiotto:2019mmf}
\begin{align}
&Z^{(1)}_{\lambda}
=
\frac{
\displaystyle (-iR_m)^{-\frac{1}{2}\sum_{i=1}^{\ell(\lambda)}\lambda_i(\lambda_i-2i+1)}
R_{y_1}^{-|\lambda|}
}{
\displaystyle 
\prod_{\Box\in\lambda}
\left(
e^{\frac{\zeta h_{\lambda}(\Box)}{2}}
-\left(-e^{-\frac{\zeta}{2}}\right)^{h_{\lambda}(\Box)}
\right)
\left(
e^{\frac{mh_{\lambda}(\Box)}{2}}
-\left(-e^{-\frac{m}{2}}\right)^{h_{\lambda}(\Box)}
\right)
}.
\label{Nf1hooklengthformula}
\end{align}
For general $N_\text{f}$, let us first reorganize the products in the integrand \eqref{ADHMgeneralNf} into the product over the variables belonging to the same Young diagram $\lambda^{(\alpha)}$ and the product of the cross-terms.
Taking into account the fact that the pole factors are always associated with either one variable $x_i$ or two variables $x_i,x_j$ belonging to the same Young diagram, we obtain
\begin{align}
Z^{(N_\text{f})}_{\lambda^{(\alpha)}}
&=
\left[
\left(
\prod_{\alpha=1}^{N_\text{f}}
\left(
\prod_{\Box\in\lambda^{(\alpha)}}\frac{1}{\displaystyle 1-((-1)^{N_\text{f}}e^{\zeta})^{h_{\lambda^{(\alpha)}}(\Box)}}\right)
\frac{1}{\left(2\cosh\frac{m}{2}\right)^{|\lambda^{(\alpha)}|}}\,
\exp\left(-\frac{i\zeta}{2\pi}\sum_{\substack{i\\ (x_i\in\lambda^{(\alpha)})}} x_i\right)\right.\right.\nonumber \\
&\quad\quad \quad \quad \left.\times \ \frac{\displaystyle {\prod}^\prime_{\substack{i<j\\ (x_i,x_j\in\lambda^{(\alpha)})}} \Bigl(2\sinh\frac{x_i-x_j}{2}\Bigr)^2}{\displaystyle 
{\prod}^\prime_{\substack{i<j,\pm\\ (x_i,x_j\in\lambda^{(\alpha)})}} 2\cosh\frac{x_i-x_j\pm m}{2}}
\frac{1}{\displaystyle 
{\prod}^\prime_{\substack{i\\ (x_i\in\lambda^{(\alpha)})}} 2\cosh\frac{x_i-y_\alpha}{2}}\right)\nonumber \\
& \times\left. \prod_{\alpha<\beta}\frac{
\displaystyle \prod_{x_i\in\lambda^{(\alpha)},x_j\in\lambda^{(\beta)}} 
\Bigl(2\sinh\frac{x_i-x_j}{2}\Bigr)^2
}{
\displaystyle \prod_{x_i\in\lambda^{(\alpha)},x_j\in\lambda^{(\beta)},\pm}2\cosh\frac{x_i-x_j\pm m}{2}}
\prod_{x_i\in\lambda^{(\alpha)}}
\frac{1}{2\cosh\frac{x_i-y_\beta}{2}}
\prod_{x_i\in\lambda^{(\beta)}}
\frac{1}{2\cosh\frac{x_i-y_\alpha}{2}}
\right]_{\text{pole}}.
\label{ZNfcombinatorial1}
\end{align}
Since the calculation of the first and second line for each $\alpha$ is essentially the same as the calculation of $Z^{(1)}_{\lambda=\lambda^{(\alpha)}}$, we find
\begin{align}
&\left[
\left(
\prod_{\Box\in\lambda^{(\alpha)}}\frac{1}{\displaystyle 1-((-1)^{N_\text{f}}e^{\zeta})^{h_{\lambda^{(\alpha)}}(\Box)}}\right)
\frac{1}{\left(2\cosh\frac{m}{2}\right)^{|\lambda^{(\alpha)}|}}\,
\exp \left(\frac{\zeta}{2\pi i}\sum_{\substack{i\\ (x_i\in\lambda^{(\alpha)})}} x_i\right)\right.\nonumber \\
&\quad \quad \quad \quad \left.\times \ \frac{\displaystyle {\prod}^\prime_{\substack{i<j\\ (x_i,x_j\in\lambda^{(\alpha)})}} \Bigl(2\sinh\frac{x_i-x_j}{2}\Bigr)^2}{\displaystyle 
{\prod}^\prime_{\substack{i<j,\pm\\ (x_i,x_j\in\lambda^{(\alpha)})}} 2\cosh\frac{x_i-x_j\pm m}{2}}
\frac{1}{\displaystyle 
{\prod}^\prime_{\substack{i\\ (x_i\in\lambda^{(\alpha)})}} 2\cosh\frac{x_i-y_\alpha}{2}}\right]_{\text{pole}}\nonumber \\
= \, &
\frac{
\displaystyle (-iR_m)^{-\frac{1}{2}\sum_{i=1}^{\ell(\lambda^{(\alpha)})}\lambda^{(\alpha)}_i(\lambda^{(\alpha)}_i-2i+1)}
R_{y_\alpha}^{-|\lambda^{(\alpha)}|}
}{
\displaystyle 
\prod_{\Box\in\lambda^{(\alpha)}}
\left(
\left((-1)^{N_\text{f}-1}e^{\frac{\zeta}{2}}\right)^{h_{\lambda^{(\alpha)}}(\Box)}
-\left(-e^{-\frac{\zeta}{2}}\right)^{h_{\lambda^{(\alpha)}}(\Box)}
\right)
\left(
e^{\frac{mh_{\lambda^{(\alpha)}}(\Box)}{2}}
-\left(-e^{-\frac{m}{2}}\right)^{h_{\lambda^{(\alpha)}}(\Box)}
\right)
}.
\label{ZNfcombinatorialdiag}
\end{align}
Here we have taken into account the fact that the quasi-periodicity factors \eqref{quasiperiodicityfactorgeneral} which determine the $\zeta$-dependence of the building block $Z_{\lambda^{(\alpha)}}^{(N_\text{f})}$ are different for even $N_\text{f}$ from those for odd $N_\text{f}$.

Next let us look into the cross terms, the factors in the third line of \eqref{ZNfcombinatorial1}.
Since the value of $x_i$ associated with a box $\Box=(a,b)\in\lambda^{(\alpha)}$ at the pole is
\begin{align}
x_i=y_\alpha+(b-a)m+\pi i(a+b-1),
\end{align}
we can write the factors in the third line of \eqref{ZNfcombinatorial1} as
\begin{align}
&\left[
\frac{
\displaystyle \prod_{x_i\in\lambda^{(\alpha)},x_j\in\lambda^{(\beta)}} 
\left(2\sinh\frac{x_i-x_j}{2}\right)^2
}{
\displaystyle \prod_{x_i\in\lambda^{(\alpha)},x_j\in\lambda^{(\beta)},\pm}2\cosh\frac{x_i-x_j\pm m}{2}}
\prod_{x_i\in\lambda^{(\alpha)}}
\frac{1}{2\cosh\frac{x_i-y_\beta}{2}}
\prod_{x_i\in\lambda^{(\beta)}}
\frac{1}{2\cosh\frac{x_i-y_\alpha}{2}}
\right]_{\text{pole}}\nonumber \\
&=g_{\lambda^{(\alpha)},\lambda^{(\beta)}}(y_\alpha-y_\beta),
\label{ZNfcombinatorialcross}
\end{align}
with
\begin{align}
&g_{\lambda,\mu}(y)
=
\left(
\prod_{\Box=(a,b)\in\lambda}\prod_{\Box'=(a',b')\in\mu}
\frac{
(2\sinh\frac{y+(b-b'-a+a')m+\pi i(a+b-a'-b')}{2})^2
}{
\prod_\pm 2\cosh\frac{y+(b-b'-a+a'\pm 1)m+\pi i(a+b-a'-b')}{2}
}
\right)\nonumber \\
&\quad\times \left(
\prod_{\Box=(a,b)\in\lambda}
\frac{1}{2\cosh\frac{y+(b-a)m+\pi i (a+b-1)}{2}}
\right)
\left(
\prod_{\Box=(a,b)\in\mu}
\frac{1}{2\cosh\frac{-y+(b-a)m+\pi i(a+b-1)}{2}}
\right).
\label{gtilde}
\end{align}
Plugging \eqref{ZNfcombinatorialdiag} and \eqref{ZNfcombinatorialcross} into \eqref{ZNfcombinatorial1}, we conclude
\begin{align}
&Z^{(N_\text{f})}_{\lambda^{(\alpha)}}=\nonumber \\
&
\left(
\prod_{\alpha=1}^{N_\text{f}}
\frac{
\displaystyle (-iR_m)^{-\frac{1}{2}\sum_{i=1}^{\ell(\lambda^{(\alpha)})}\lambda_i^{(\alpha)}(\lambda_i^{(\alpha)}-2i+1)}
R_{y_\alpha}^{-|\lambda^{(\alpha)}|}
}{
\displaystyle 
\prod_{\Box\in\lambda^{(\alpha)}}
\left(
\left((-1)^{N_\text{f}-1}e^{\frac{\zeta}{2}}\right)^{h_{\lambda^{(\alpha)}}(\Box)}
-(-e^{-\frac{\zeta}{2}})^{h_{\lambda^{(\alpha)}}(\Box)}
\right)
\left(
e^{\frac{mh_{\lambda^{(\alpha)}}(\Box)}{2}}
-\left(-e^{-\frac{m}{2}}\right)^{h_{\lambda^{(\alpha)}}(\Box)}
\right)
}\right)\nonumber \\
& \ \quad\times \prod_{\alpha<\beta}g_{\lambda^{(\alpha)},\lambda^{(\beta)}}(y_\alpha-y_\beta).
\label{ZNfcombinatorial2}
\end{align}

When the Young diagram is not of hook type, the correspondence between the Young diagram and the set of the pole factors according to the rule displayed in the previous subsection is not in one-to-one.
As a result we entounter the double poles, which are compensated with the zeroes in the numerator.
Although the calculation involving non-hook Young diagrams is subtle, in \cite{Gaiotto:2019mmf} it was claimed for $N_\text{f}=1$ that the formula \eqref{Nf1hooklengthformula} still holds for these cases.
Indeed, the exact values of the partition function for $N\ge 4$ obtained by the combinatorial formula \eqref{totalZincombinatorialformula} with \eqref{Nf1hooklengthformula} coincide with the exact values calculated by the Fermi gas formalism \cite{Hatsuda:2012hm,Honda:2018pqa,Nosaka:2020tyv}.
For this reason, we claim that our final formula \eqref{ZNfcombinatorial2} for the building block $Z^{(N_\text{f})}_{\lambda^{(\alpha)}}$ is also valid for any set of Young diagrams $(\lambda^{(1)},\cdots,\lambda^{(N_\text{f})})$, even when it contains non-hook diagrams.
In section \ref{sec_evidenceforcombinatorialformula} we provide another non-trivial check for the combinatorial formula \eqref{ZNfcombinatorial2} by using the bilinear equations for the grand canonical partition functions.

Note that in $g_{\lambda,\mu}(y)$ which gives the cross term, all sinh factors in the numerators are cancelled with a part of the denominator.
As a result we can rewrite $g_{\lambda,\mu}(y)$ in a more compact form as
\begin{align}
g_{\lambda,\mu}(y)
=
E_{\lambda,\mu}(y)
\times
\prod_{\Box\in\lambda}e^{-\frac{\pi i(\text{arm}_\lambda(\Box)-\text{leg}_\mu(\Box)}{2}}
\prod_{\Box\in\mu}e^{\frac{\pi i(\text{arm}_\mu(\Box)-\text{leg}_\lambda(\Box)}{2}}.
\label{gtildevsNekrasov}
\end{align}
with
\begin{align}
&E_{\lambda,\mu}(y)=
\frac{1}{\displaystyle \prod_{\Box\in\lambda}
\left(e^{\frac{y}{2}+\frac{m}{2}(\text{arm}_\lambda(\Box)+\text{leg}_\mu(\Box)+1)}+e^{-\frac{y}{2}-\frac{m}{2}(\text{arm}_\lambda(\Box)+\text{leg}_\mu(\Box)+1)-\pi i(\text{arm}_\lambda(\Box)-\text{leg}_\mu(\Box))}\right)
}\nonumber \\
&\quad\quad\times \frac{1}{\displaystyle \prod_{\Box\in\mu}
\left(e^{\frac{y}{2}-\frac{m}{2}(\text{arm}_\mu(\Box)+\text{leg}_\lambda(\Box)+1)}+e^{-\frac{y}{2}+\frac{m}{2}(\text{arm}_\mu(\Box)+\text{leg}_\lambda(\Box)+1)+\pi i(\text{arm}_\mu(\Box)-\text{leg}_\lambda(\Box))}\right)
}.
\label{Nf2combinatorialformula_Elambdamuagain}
\end{align}
Here $\text{arm}_\lambda(\Box)$ and $\text{leg}_\lambda(\Box)$ are the arm/leg length of the box $\Box$ defined as \eqref{armleg1},\eqref{armleg2}.
See appendix \ref{app_proofofgtildevsNekrasov} for the proof of the identity \eqref{gtildevsNekrasov}.

\section{Bilinear equations}
\label{sec_bilineq}

\subsection{$N_\text{f}=1$ (review)}

Before explaining our new results for $N_\text{f}\ge 2$, let us first review in detail the bilinear relation for the $N_\text{f}=1$ matrix model found in \cite{Nosaka:2024gle}, where some method we will use to guess the bilinear equations for $N_\text{f}\ge 2$ were already exploited.
In \cite{Grassi:2014uua,Bonelli:2017gdk} it was proposed that the grand canonical partition function of the $\text{U}(N)_k\times \text{U}(N+M)_{-k}$ ABJ satisfies the bilinear difference equation with respect to $M$, which is of the form of the $\mathfrak{q}$-deformed Painlev\'e $\text{III}_3$ equation.
Although the bilinear equations are originally verified for $k\ge 2$ and $M=1,2,\cdots,k-1$ so that the relative ranks $M'$ appearing in the bilinear equations are restricted to the range $0\le M'\le k$, in \cite{Moriyama:2023pxd} the equation was extended to $M=0$ (and $M=k$) by relating the grand partition function at $M=-1$ (and $M=k+1$) with the grand partition at $M=k-1$ (and $M=1$).
In particular, after this extension the bilinear equation exists also for $k=1$.

\begin{figure}[tp]
\centering
\includegraphics[width=\textwidth]{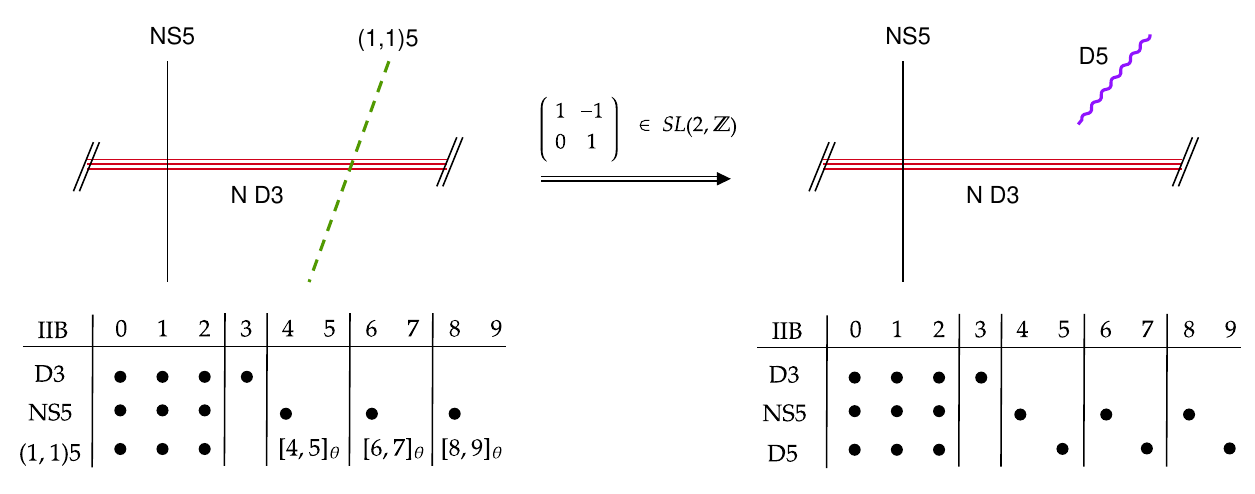}
\caption{Illustration of the duality between ABJM theory with $k=1$ and 3d $\mathcal{N}=4$ Yang-Mills theory with one flavor.
The $(1,1)5$-brane extended with the angle $\theta = \arctan 1=\pi/4$ transforms,  under SL(2,$\mathbb{Z}$) transformation of the type IIB string theory, to a D5-brane on the right hand side.
}
\label{duality}
\end{figure}

Using the duality between the ABJM theory with the Chern-Simons level $k=1$ and the 3d ${\cal N}=4$ Yang-Mills theory with one adjoint hypermultiplet and one fundamental hypermultiplet, namely \eqref{ADHMgeneralNf} with $N_\text{f}=1$ and $y_1=0$ (see figure \ref{duality} for the brane setup), this equation can be written as the equation for the grand partition function of the super Yang-Mills theory 
\begin{align}
\Xi^{(1)}_0(\zeta,m;u)=\sum_{N=0}^\infty u^NZ_{0}^{(1)}(\zeta,m;N)
\label{XiofNf1}
\end{align}
as
\begin{align}
\Xi^{(1)}_0(0,0;u)^2-\Xi^{(1)}_0(0,0;-u)^2-u\prod_\pm \Xi^{(1)}_0(0,0;\pm iu)=0.
\end{align}

Motivated by these results, it was proposed in \cite{Nosaka:2024gle} that the $\text{U}(N)_k\times \text{U}(N+M)_{-k}$ ABJ theory with two-parameter mass deformations also satisfy the bilinear equations, which in particular for $k=1,M=0$ and written in terms of the grand partition function of the super Yang-Mills theory \eqref{XiofNf1} takes the following form
\begin{align}
a_1 \prod_\pm \Xi_0^{(1)}(b_1^{\pm 1}u)
+a_2 \prod_\pm \Xi_0^{(1)}(b_2^{\pm 1}u)
+cu \prod_\pm \Xi_0^{(1)}(d^{\pm 1}u)=0.
\label{bilinNf1withmzetaguess}
\end{align}
Here and afterward we abbreviate $Z^{(N_\text{f})}_{y_\alpha}(\zeta,m,N)$ and $\Xi^{(N_\text{f})}_{y_\alpha}(\zeta,m;u)$ respectively as $Z^{(N_\text{f})}_{y_\alpha}(N)$ and $\Xi^{(N_\text{f})}_{y_\alpha}(u)$.
The coefficients $a_1,a_2,b_1,b_2,c,d$ can be guessed from the exact values of the partition function for $N\le 2$ \eqref{totalZincombinatorialformula} with \eqref{Nf1hooklengthformula}
\begin{align}
Z_0^{(1)}(0)=1,\quad
Z_0^{(1)}(1)=\frac{1}{2\cosh\frac{\zeta}{2} \, 2\cosh\frac{m}{2}},\quad
Z_0^{(1)}(2)=\frac{i(R_m^{-1}-R_m)}{2\sinh\zeta \, 2\cosh\frac{\zeta}{2} \, 2\sinh m \, 2\cosh\frac{m}{2}}
\label{exactvaluesNf0N012}
\end{align}
with $R_m=e^{\frac{i\zeta m}{2\pi}}$ \eqref{RmRy}.
Indeed, by setting $u=0$ in bilinear equation \eqref{bilinNf1withmzetaguess} we obtain the equation which determines $a_1,a_2$ as (up to the overall rescaling of the bilinear equation)
\begin{align}
a_1=1,\quad a_2=-1,
\end{align}
while by expanding the bilinear equation in $u$ and looking at the coefficients of $u$ and $u^2$ we obtain the equations for $Z_0^{(1)}(1)$ and $Z_0^{(1)}(2)$ as
\begin{align}
Z_0^{(1)}(1)=-\frac{cZ_0^{(1)}(0)}{b_1+b_1^{-1}-b_2-b_2^{-1}},\quad
Z_0^{(1)}(2)=-\frac{c(d+d^{-1})Z_0^{(1)}(1)}{b_1^2+b_1^{-2}-b_2^2-b_2^{-2}},
\label{Nf1ZN2frombilineq}
\end{align}
where we have also used $a_1=-a_2=1$.
By comparing these equation with the exact values \eqref{exactvaluesNf0N012} written as
\begin{align}
Z_0^{(1)}(1)=\frac{1}{
e^{\frac{\zeta+m}{2}}
+e^{-\frac{\zeta+m}{2}}
+e^{-\frac{\zeta-m}{2}}
+e^{\frac{\zeta-m}{2}}
},\quad
Z_0^{(1)}(2)=\frac{i(R_m^{-1}-R_m)Z_0^{(1)}(1)}{2\sinh\zeta \, 2\sinh m},
\end{align}
it is not difficult to guess the coefficients $b_1,b_2,c,d$ as
\begin{align}
b_1=e^{\frac{\zeta+m}{2}},\quad
b_2=-e^{-\frac{\zeta-m}{2}},\quad
c=-1,\quad
d=-iR_m.
\end{align}
Hence the bilinear equation is guessed as
\begin{align}
\prod_\pm \Xi^{(1)}_0\left(e^{\pm\frac{\zeta+m}{2}}u\right)
-\prod_\pm \Xi^{(1)}_0\left(-e^{\mp\frac{\zeta-m}{2}}u\right)
-u\prod_\pm \Xi^{(1)}_0\left(\mp iR_m^{\pm 1}u\right)=0.
\label{bilineqNf1withzetamguessed}
\end{align}

Similarly, expanding the bilinear equation \eqref{bilineqNf1withzetamguessed} and looking at the coefficients of $u^N$, we obtain the recursion relation for the exact partition function $Z^{(1)}_0(N)$ with respect to $N$
\begin{align}
&\quad \quad \quad \quad Z^{(1)}_0(N)=\frac{1}{2\cosh\frac{(\zeta+m)N}{2}-(-1)^N2\cosh\frac{(\zeta-m)N}{2}}
\nonumber \\
&\times\left[
-\sum_{N'=1}^{N-1}(e^{\frac{(\zeta+m)(2N'-N)}{2}}-(-1)^Ne^{-\frac{(\zeta-m)(2N'-N)}{2}})Z^{(1)}_0(N')Z^{(1)}_0(N-N')\right.\nonumber \\
&\quad \quad \quad \left.+\sum_{N'=0}^{N-1}(-iR_m)^{2N'-N+1}Z^{(1)}_0(N')Z^{(1)}_0(N-1-N')
\right],
\end{align}
which gives the exact values of the partition function also for $N\ge 3$, which are consistent with the exact values obtained by direct calculations \cite{Gaiotto:2019mmf,Nosaka:2020tyv}.
In particular, from the recursion relation it follows that the partition function is expanded in $R_m$ as
\begin{align}
Z^{(1)}_0(N)=\!\!\!\!\!\!\!\sum_{a=-\frac{N(N-1)}{2}}^{\frac{N(N-1)}{2}}\!\!\!\!\!\! R_m^af_a(\zeta,m),
\end{align}
with $f_a(\zeta,m)$ some rational functions of $e^{\frac{\zeta}{2}}$ and $e^{\frac{m}{2}}$.
This range of the powers of $R_m$ is in agreement with the combinatorial formula \eqref{totalZincombinatorialformula} with \eqref{Nf1hooklengthformula}, where the highest/lowest power of $R_m$ in $Z^{(1)}_0(N)$ is realized by the Young diagrams $\lambda=(1^N)$ and $\lambda=(N)$.

In the next subsection we generalize the bilinear equation \eqref{bilineqNf1withzetamguessed} to the super Yang-Mills theory with general number $N_\text{f}$ of hypermultiplets by following the same strategy.

\subsection{$N_\text{f}=2$ and higher}

First let us consider the case with two flavors.
Without loss of generality, we set $y_1=-y_2=y$.
We also denote $R_{y_1}=R_{y_2}^{-1}$ as $R_y$ and $Z_{y_\alpha}^{(2)}(N)$ as $Z_y^{(2)}(N)$.
The exact values of the partition function for $N=1,2$ are obtained from \eqref{totalZincombinatorialformula} with \eqref{ZNfcombinatorial2} as $Z^{(2)}_y(0)=1$ and
\begin{subequations}
\begin{align}
&Z_y^{(2)}(1)=\frac{i(R_y^{-1}-R_y)}{(2\sinh\frac{\zeta}{2})(2\cosh\frac{m}{2})(2\sinh y)},\label{Nf2exactvaluesforN1} \\
&Z_y^{(2)}(2)=\nonumber \\
&\sum_\pm \frac{\pm R_m^{\mp 1}}{\left(2\sinh\zeta\right)(2\sinh\frac{\zeta}{2})(2\sinh m)(2\cosh\frac{m}{2})(2\sinh y)}\left(\frac{R_y^{-2}}{2\cosh(y\pm\frac{m}{2})}-\frac{R_y^2}{2\cosh(y\mp\frac{m}{2})}\right)\nonumber \\
&\quad\quad +\frac{1}{(2\sinh\frac{\zeta}{2})^2(2\cosh\frac{m}{2})^2\prod_\pm 2\cosh(y\pm \frac{m}{2})}.
\label{Nf2exactvaluesforN2}
\end{align}
\label{Nf2exactvaluesforN1N2}
\end{subequations}

Now let us guess the bilinear equation for the grand partition function
\begin{align}
\Xi^{(2)}_{y}(u)=\sum_{N=0}^\infty u^NZ_y^{(2)}(N)
\end{align}
by using the exact values \eqref{Nf2exactvaluesforN1N2}.
From the exact value at $N=1$, let us first suppose that the bilinear equation takes the form
\begin{align}
\sum_{i=1}^{L_1}a_i\prod_\pm \Xi_y(b_i^{\pm 1}u)+uR_y\sum_{i=1}^{L_2}c_i\prod_\pm\Xi_y(d_i^{\pm 1}u)
+uR_y^{-1}\sum_{i=1}^{L_2}c_i'\prod_\pm\Xi_y(d_i'{}^{\pm 1}u)
\mathop{=}^?0.
\label{bilineqdoesnotwork}
\end{align}
with $L_1,L_2$ some positive integers and $a_i,b_i,c_i,c'_i$ some functions of $\zeta,m,y$ which do not contain $R_m$ and $R_y$.
In the combinatorial formula for $N_\text{f}=2$ the highest power of $R_m$ in $Z^{(2)}_y(N)$ is $R_m^{\frac{N(N-1)}{2}}$.
Hence as we have argued above for $N_\text{f}=1$ by using the recursion relation obtained from the bilinear equation, at least one of $d_i,d_i{}'$ should contain some powers of $R_m$.
On the other hand, the highest power of $R_y$ in $Z^{(2)}_y(N)$ is only linear in $N$, $R_y^N$, which suggests that none of $d_i,d_i{}'$ should contain the powers of $R_y$.

However, we find that this ansatz \eqref{bilineqdoesnotwork} for the bilinear equation does not work.
As we have demonstrated above for $N_\text{f}=1$ \eqref{Nf1ZN2frombilineq}, if \eqref{bilineqdoesnotwork} was satisfied the partition function at $N=2$ would be proportional to $Z^{(2)}_y(1)$ with a simple overall factor.
On the other hand, the actual exact values $Z^{(2)}_y(2)$ \eqref{Nf2exactvaluesforN2} does not satisfy such relation.

In order to find a new ansatz for the bilinear equation, let us focus on the powers of $R_m$ and $R_y$ which appear in $Z^{(2)}_y(2)$, $R_mR_y^2$, $R_mR_y^{-2}$, $R_m^{-1}R_y^2$, $R_m^{-1}R_y^{-2}$ and $R_m^0R_y^0$.
We notice that all these five powers are associated with the partition function at $N=1$ with shifted $y$-variable:
\begin{itemize}
\item $R_yR_m^{\frac{1}{2}}Z^{(2)}_{y+\frac{m}{2}}(1)$ contains $R_m^0R_y^0$ and $R_mR_y^2$,
\item $R_yR_m^{-\frac{1}{2}}Z^{(2)}_{y-\frac{m}{2}}(1)$ contains $R_m^0R_y^0$ and $R_m^{-1}R_y^2$,
\item $R_y^{-1}R_m^{-\frac{1}{2}}Z^{(2)}_{y+\frac{m}{2}}(1)$ contains $R_m^{-1}R_y^{-2}$ and $R_m^0R_y^0$,
\item $R_y^{-1}R_m^{\frac{1}{2}}Z^{(2)}_{y-\frac{m}{2}}(1)$ contains $R_mR_y^{-2}$ and $R_m^0R_y^0$.
\end{itemize}
Remarkably, these four terms do not contain any unwanted powers of $R_m$ and $R_y$.
From this observation, let us suppose that the grand partition function for $N_\text{f}=2$ satisfies the bilinear equation of the following form:
\begin{align}
&\sum_{i=1}^{L_1}a_i\prod_\pm \Xi_y\left(b_i^{\pm 1}u\right)
+uR_y\sum_{i=1}^{L_2}c_i\prod_\pm\Xi_{y\pm(\frac{m}{2}+\Delta_i)}\left((R_m^{\frac{1}{2}}\tilde{d}_i)^{\pm 1}u\right)\nonumber \\
&\quad\quad +uR_y^{-1}\sum_{i=1}^{L_2}c_i'\prod_\pm\Xi_{y\pm(\frac{m}{2}+\Delta'_i)}\left((R_m^{-\frac{1}{2}}\tilde{d}_i')^{\pm 1}u\right)
=0,
\label{bilineqNf2newansatz}
\end{align}
with $L_1,L_2$ some positive integers, $a_i,b_i,c_i,c'_i,\tilde{d}_i,\tilde{d}_i'$ some functions of $\zeta,m,y$ which do not contain $R_m$ and $R_y$, and $\Delta_i,\Delta_i'$ some constants independent of $\zeta,m,y$.
Note that similar shift of the continuous parameters also appeared in the bilinear equation for the grand partition function of the four-node circular quiver super Chern-Simons threoy with levels $k,0,-k,0$ \cite{Bonelli:2022dse,Moriyama:2023mjx}.
The bilinear equations for this quiver Chern-Simons theory is the $\mathfrak{q}$-Painlev\'e VI equation in $\tau$-form \cite{Jimbo:2017ael} involving the shift of the discrete time variable and the four Painlev\'e parameters, which are mapped to the three relative ranks and two FI parameters of the super Chern-Simons theory.
In the type IIB brane construction the two FI parameters corresponds to the relative positions of the 5-branes of the same kind.
In particular, for $k=1$ these parameters are indeed naturally related through the $\text{SL}(2,\mathbb{Z})$ transformation of the brane setup to the masses of the fundamental hypermultiplets in the dual theory.

First let us look at the equations obtained from the coefficients of $u^0$ and $u^1$ of the bilinear equation \eqref{bilineqNf2newansatz}, where the shifts of $y$ do not play any role.
\begin{align}
\sum_{i=1}^{L_1}a_i=0,\quad
Z^{(2)}_y(1)=-\frac{R_y\sum_{i=1}^{L_2}c_i+R_y^{-1}\sum_{i=1}^{L_2}c_i'}{\sum_{i=1}^{L_1}a_i(b_i+b_i^{-1})}.
\end{align}
By comparing these with the exact value of $Z^{(2)}_y(1)$ \eqref{Nf2exactvaluesforN1}, we find the following simple choice of the coefficients $a_i,b_i,c_i,c'_i$ for $L_1=2$ and $L_2=1$
\begin{align}
a_1=1,\quad
a_2=-1,\quad
b_1=ie^{\frac{\zeta+m}{2}},\quad
b_2=ie^{-\frac{\zeta-m}{2}},\quad
c_1=-\frac{1}{2\sinh y},\quad
c_1'=\frac{1}{2\sinh y}.
\label{Nf2abcc'}
\end{align}
Now let us look at the equation obtained from the coefficient of $u^2$ of the bilinear equation \eqref{bilineqNf2newansatz}.
By choosing $L_1=2,L_2=1$ and substituting the coefficients $a_i,b_i,c_1,c_1'$ \eqref{Nf2abcc'}, we obtain
\begin{align}
Z^{(2)}_y(2)&=\frac{1}{2\sinh\zeta\, 2\sinh m \, 2\sinh y}\biggl[
-R_y\Bigl(
R_m^{\frac{1}{2}}\tilde{d}_1Z_{y+\frac{m}{2}+\Delta_1}^{(2)}(1)
+R_m^{-\frac{1}{2}}\tilde{d}_1^{-1}Z_{y-\frac{m}{2}-\Delta_1}^{(2)}(1)
\Bigr)\nonumber \\
&\quad +R_y^{-1}\Bigl(
R_m^{-\frac{1}{2}}\tilde{d}_1'Z_{y+\frac{m}{2}+\Delta_1'}^{(2)}(1)
+R_m^{\frac{1}{2}}\tilde{d}_1'{}^{-1}Z_{y-\frac{m}{2}-\Delta_1'}^{(2)}(1)
\Bigr)
\biggr]\nonumber \\
&=
\frac{i\tilde{d}_1e^{\frac{i\zeta\Delta_1}{2\pi}}R_mR_y^2}{2\sinh\zeta\, 2\sinh\frac{\zeta}{2} \, 2\sinh m \, 2\cosh\frac{m}{2} \, 2\sinh y \, 2\sinh(y+\frac{m}{2}+\Delta_1)}\nonumber \\
&\quad +\frac{i\tilde{d}'_1{}^{-1}e^{\frac{i\zeta\Delta_1'}{2\pi}}R_mR_y^{-2}}{2\sinh\zeta \, 2\sinh\frac{\zeta}{2} \, 2\sinh m \, 2\cosh\frac{m}{2} \, 2\sinh y \, 2\sinh(y-\frac{m}{2}-\Delta_1')}\nonumber \\
&\quad +\frac{i\tilde{d}_1{}^{-1}e^{-\frac{i\zeta\Delta_1}{2\pi}}R_m^{-1}R_y^2}{2\sinh\zeta \, 2\sinh\frac{\zeta}{2} \, 2\sinh m \, 2\cosh\frac{m}{2} \, 2\sinh y \, 2\sinh(y-\frac{m}{2}-\Delta_1)}\nonumber \\
&\quad +\frac{i\tilde{d}_1e^{-\frac{i\zeta\Delta_1'}{2\pi}}R_m^{-1}R_y^{-2}}{2\sinh\zeta \, 2\sinh\frac{\zeta}{2}2\sinh m \, 2\cosh\frac{m}{2} \, 2\sinh y \, 2\sinh(y+\frac{m}{2}+\Delta_1')}\nonumber \\
&\quad +\frac{1}{2\sinh\zeta 2\sinh\frac{\zeta}{2} \, 2\sinh m \, 2\cosh\frac{m}{2} \, 2\sinh y} \left[
-\frac{i\tilde{d}_1e^{-\frac{i\zeta\Delta_1}{2\pi}}}{2\sinh(y+\frac{m}{2}+\Delta_1)}\right.\nonumber \\
&\quad \left.-\frac{i\tilde{d}_1^{-1}e^{\frac{i\zeta\Delta_1}{2\pi}}}{2\sinh(y-\frac{m}{2}-\Delta_1)}
-\frac{i\tilde{d}_1'e^{\frac{i\zeta\Delta_1'}{2\pi}}}{2\sinh(y+\frac{m}{2}+\Delta_1')}
-\frac{i\tilde{d}_1'{}^{-1}e^{-\frac{i\zeta\Delta_1'}{2\pi}}}{2\sinh(y-\frac{m}{2}-\Delta_1')}
\right].
\label{ZNf2N2frombilineq}
\end{align}
Here in the second expression we have calculated the terms proportional respectively to $R_mR_y^2$, $R_mR_y^{-2}$, $R_m^{-1}R_y^2$, $R_m^{-1}R_y^{-2}$ and $R_m^0R_y^0$ explicitlly by using $Z^{(2)}_y(1)$ \eqref{Nf2exactvaluesforN1}.
Note that the first four terms are contributed only by the first term, fourth term, second term and third term in the first expression of \eqref{ZNf2N2frombilineq} respectively.
By comparing the first four terms with those in the actual exact value of $Z^{(2)}_y(2)$ \eqref{Nf2exactvaluesforN2}, we obtain the following equations for $\tilde{d}_1,\tilde{d}_1',\Delta_1,\Delta_1'$
\begin{align}
&\frac{1}{2\cosh(y+\frac{m}{2})}=\frac{i\tilde{d}_1e^{\frac{i\zeta \Delta_1}{2\pi}}}{2\sinh(y+\frac{m}{2}+\Delta_1)},\quad
-\frac{1}{2\cosh(y-\frac{m}{2})}=\frac{i\tilde{d}_1'{}^{-1}e^{\frac{i\zeta \Delta_1'}{2\pi}}}{2\sinh(y-\frac{m}{2}-\Delta_1')},\nonumber \\
&-\frac{1}{2\cosh(y-\frac{m}{2})}=\frac{i\tilde{d}_1^{-1}e^{-\frac{i\zeta \Delta_1}{2\pi}}}{2\sinh(y-\frac{m}{2}-\Delta_1)},\quad
\frac{1}{2\cosh(y+\frac{m}{2})}=\frac{i\tilde{d}_1'{}e^{-\frac{i\zeta \Delta_1'}{2\pi}}}{2\sinh(y+\frac{m}{2}+\Delta_1')},
\end{align}
which are solved for example by
\begin{align}
\tilde{d}_1=\sigma e^{\frac{\zeta\sigma}{4}},\quad
\tilde{d}_1'=\sigma' e^{-\frac{\zeta'\sigma}{4}},\quad
\Delta_1=\frac{\pi i\sigma}{2},\quad
\Delta_1'=\frac{\pi i\sigma'}{2},\quad (\sigma,\sigma'=\pm1).
\end{align}
Substituting these to the term proportional to $R_m^0R_y^0$ in \eqref{ZNf2N2frombilineq} and comparing it with the exact value \eqref{Nf2exactvaluesforN2}, we obtain the following equation
\begin{align}
\frac{1}{\prod_\pm 2\cosh(y\pm\frac{m}{2})}
=
\frac{1}{2\cosh\frac{\zeta}{2} \, 2\sinh \frac{m}{2} \, 2\sinh y}\left(
-\frac{e^{\frac{\zeta\sigma}{2}}+e^{-\frac{\zeta\sigma'}{2}}}{2\cosh(y+\frac{m}{2})}
+\frac{e^{-\frac{\zeta\sigma}{2}}+e^{\frac{\zeta\sigma'}{2}}}{2\cosh(y-\frac{m}{2})}
\right),
\end{align}
which is satisfied if we choose $\sigma,\sigma'$ as $\sigma'=\sigma$.
In summary, we have found the following two bilinear equations
\begin{subequations}
\begin{align}
&\prod_\pm \Xi^{(2)}_{y}\left(\pm ie^{\pm \frac{\zeta+m}{2}}u\right)-\prod_\pm\Xi^{(2)}_{y}\left(\pm ie^{\mp\frac{\zeta-m}{2}}u\right)
-\frac{uR_y}{2\sinh y}\prod_\pm \Xi^{(2)}_{y\pm(\frac{m}{2}+\frac{\pi i}{2})}\left(e^{\pm \frac{\zeta}{4}}R_m^{\pm \frac{1}{2}}u\right)\nonumber \\
&\quad\quad\quad\quad\quad\quad\quad\quad\quad\quad\quad\quad\quad\quad\quad\quad\quad\quad +\frac{uR_y^{-1}}{2\sinh y}\prod_\pm \Xi^{(2)}_{y\mp(\frac{m}{2}+\frac{\pi i}{2})}\left(e^{\pm \frac{\zeta}{4}}R_m^{\pm \frac{1}{2}}u\right)
=0,\label{ADHMNf2bilineqmzetaindepsignchoice1} \\
&\prod_\pm \Xi^{(2)}_{y}\left(\pm ie^{\pm \frac{\zeta+m}{2}}u\right)-\prod_\pm\Xi^{(2)}_{y}\left(\pm ie^{\mp\frac{\zeta-m}{2}}u\right)
-\frac{uR_y}{2\sinh y}\prod_\pm \Xi^{(2)}_{y\pm(\frac{m}{2}-\frac{\pi i}{2})}\left(-e^{\mp \frac{\zeta}{4}}R_m^{\pm \frac{1}{2}}u\right)\nonumber \\
&\quad\quad\quad\quad\quad\quad\quad\quad\quad\quad\quad\quad\quad\quad\quad\quad\quad\quad +\frac{uR_y^{-1}}{2\sinh y}\prod_\pm \Xi^{(2)}_{y\mp(\frac{m}{2}-\frac{\pi i}{2})}\left(-e^{\mp \frac{\zeta}{4}}R_m^{\pm \frac{1}{2}}u\right)
=0,
\label{ADHMNf2bilineqmzetaindepsignchoice2}
\end{align}
\end{subequations}
which are consistent with the exact values of $Z^{(2)}_y(N)$ at least for $N=0,1,2$.

Note that the two bilinear equations are related by a change of variable $m\rightarrow -m$.
To see this let us denote the $m$-dependence of the partition function and the grand partition function explicitly as $\Xi_{y}^{(2)}(m;u)=\sum_{N=0}^\infty u^NZ^{(2)}_{y}(m;N)$.
Then, by setting $m=-m'$ and $u=-u'$ in the second bilinear equation \eqref{ADHMNf2bilineqmzetaindepsignchoice2} we obtain
\begin{align}
&-\prod_\pm \Xi^{(2)}_{y}\left(-m';\pm ie^{\pm \frac{\zeta+m'}{2}}u'\right)+\prod_\pm\Xi^{(2)}_{y}\left(-m';\pm ie^{\mp\frac{\zeta-m'}{2}}u'\right)\nonumber \\
&+\frac{u'R_y}{2\sinh y}\prod_\pm \Xi^{(2)}_{y\pm(\frac{m'}{2}+\frac{\pi i}{2})}\!\!\left(\!-m';e^{\pm \frac{\zeta}{4}}R_{m'}^{\pm \frac{1}{2}}u'\!\right)\!\!
-\frac{u'R_y^{-1}}{2\sinh y}\prod_\pm \Xi^{(2)}_{y\mp(\frac{m'}{2}+\frac{\pi i}{2})}\!\!\left(\!-m';e^{\pm \frac{\zeta}{4}}R_{m'}^{\pm \frac{1}{2}}u'\!\right)\!\!
=0,
\end{align}
where $R_{m'}=e^{\frac{i\zeta m'}{2\pi}}$.
This equation is identical to the first bilinear equation \eqref{ADHMNf2bilineqmzetaindepsignchoice1} with $m$ replaced with $m'$, except the argument $-m'$ of the grand partition functions.
Namely, the second equation \eqref{ADHMNf2bilineqmzetaindepsignchoice2} simply says that the grand partition function defined from $Z_{y}^{(2)}(-m;N)$ instead of $Z_{y}^{(2)}(m;N)$ also satisfies the same equation \eqref{ADHMNf2bilineqmzetaindepsignchoice1} as the original grand partition function.
Since in our setup the partition function $Z_{y}^{(2)}(m;N)$ \eqref{ADHMgeneralNf} is manifestly symmetric under $m\rightarrow -m$, the second bilinear equation \eqref{ADHMNf2bilineqmzetaindepsignchoice2} does not tell us any new properties of the partition function.

Once we realize the necessity of the shifts of the $y_\alpha$-variables, it is straightforward to repeat the same analysis to find the bilinear equations satisfied by the grand partition function for general $N_\text{f}$
\begin{align}
\Xi_{y_\alpha}^{(N_\text{f})}(u)=\sum_{N=0}^\infty u^NZ_{y_\alpha}^{(N_\text{f})}(N).
\end{align}
As a result, for even $N_\text{f}$ we find\footnote{
Here and in the remaining equations of this subsection, we write the short-hand subscript $y_\alpha$ explicitly in order to show the shift structures on them.
}
\begin{align}
&\prod_\pm \Xi_{y_1,\cdots,y_{N_\text{f}}}^{(N_\text{f})}\left(\pm ie^{\pm \frac{\zeta+m}{2}}u\right)
-\prod_\pm \Xi_{y_1,\cdots,y_{N_\text{f}}}^{(N_\text{f})}\left(\pm ie^{\mp \frac{\zeta-m}{2}}u\right)\nonumber \\
&\quad
-i^{N_\text{f}}
u\sum_{\beta=1}^{N_\text{f}}\frac{R_{y_\beta}^{-1}}{\prod_{\gamma(\neq\beta)}2\sinh\frac{y_\beta-y_\gamma}{2}}
\prod_\pm \Xi_{y_1\pm\Delta^{(\beta)}_1,\cdots,y_{N_\text{f}}\pm\Delta^{(\beta)}_{N_\text{f}}}^{(N_\text{f})}\left(R_m^{\pm \frac{1}{N_\text{f}}}e^{\pm\frac{1}{2}(1-\frac{1}{N_\text{f}})\zeta}u\right)=0,
\label{generalevenNf}
\end{align}
with
\begin{align}
\Delta^{(\beta)}_\alpha=\begin{cases}
\displaystyle \Bigl(\frac{1}{N_\text{f}}-1\Bigr)(m+\pi i)&\quad (\alpha=\beta)\vspace{0.2cm} \\
\displaystyle \frac{m+\pi i}{N_\text{f}}&\quad (\alpha\neq\beta)
\end{cases}.
\label{Deltaalphabeta}
\end{align}
For $N_\text{f}=2$, the bilinear equation \eqref{generalevenNf} reduces to the equation \eqref{ADHMNf2bilineqmzetaindepsignchoice1} we have found above.

For odd $N_\text{f}$, we find
\begin{align}
&\prod_\pm \Xi_{y_1,\cdots,y_{N_\text{f}}}^{(N_\text{f})}\left(e^{\pm \frac{\zeta+m}{2}}u\right)
-\prod_\pm \Xi_{y_1,\cdots,y_{N_\text{f}}}^{(N_\text{f})}\left(-e^{\mp \frac{\zeta-m}{2}}u\right)\nonumber \\
&\quad -i^{N_\text{f}-1}
u\sum_{\beta=1}^{N_\text{f}}\frac{R_{y_\beta}^{-1}}{\prod_{\gamma(\neq\beta)}2\sinh\frac{y_\beta-y_\gamma}{2}}
\prod_\pm \Xi_{y_1\pm\Delta^{(\beta)}_1,\cdots,y_{N_\text{f}}\pm\Delta^{(\beta)}_{N_\text{f}}}^{(N_\text{f})}\left(\mp iR_m^{\pm \frac{1}{N_\text{f}}}e^{\pm\frac{1}{2}\left(1-\frac{1}{N_\text{f}}\right)\zeta}u\right)=0 ,
\label{generaloddNf}
\end{align}
with $\Delta^{(\beta)}_\alpha$ given as \eqref{Deltaalphabeta}.
For $N_\text{f}=1$, the bilinear equation \eqref{generaloddNf} reduces to the equation \eqref{bilineqNf1withzetamguessed} for the mass deformed ABJM theory.

From the coefficients of $u^N$ of the bilinear equation \eqref{generalevenNf} and \eqref{generaloddNf}, we obtain the following recursion relation for $Z^{(N_\text{f})}_{y_\alpha}(N)$ in $N$
\begin{align}
&Z_{y_1,\cdots,y_{N_\text{f}}}^{(N_\text{f})}(N)=\frac{1}{2\cosh\frac{(\zeta+m+\pi i)N}{2}-2\cosh\frac{(\zeta-m-\pi i)N}{2}} \nonumber \\
&\quad \times \left[ -\sum_{N'=1}^{N-1}\left(e^{\frac{(\zeta+m+\pi i)(2N'-N)}{2}}
-e^{-\frac{(\zeta-m-\pi i)(2N'-N)}{2}}
\right)Z_{y_1,\cdots,y_{N_\text{f}}}^{(N_\text{f})}(N')Z_{y_1,\cdots,y_{N_\text{f}}}^{(N_\text{f})}(N-N')\right.\nonumber \\
&\quad \quad \quad +
i^{N_\text{f}}\sum_{\beta=1}^{N_\text{f}}\frac{R_{y_\beta}^{-1}}{\prod_{\gamma(\neq\beta)}2\sinh\frac{y_\beta-y_\gamma}{2}}\sum_{N'=0}^{N-1}\left(R_m^{\frac{1}{N_\text{f}}}e^{\frac{1}{2}(1-\frac{1}{N_\text{f}})\zeta}\right)^{2N'-N+1}\nonumber \\
&\quad\quad\quad\times \left.Z^{(N_{\text{f}})}_{y_1+\Delta^{(\beta)}_1,\cdots,y_{N_\text{f}}+\Delta^{(\beta)}_{N_\text{f}}}(N')
Z^{(N_{\text{f}})}_{y_1-\Delta^{(\beta)}_1,\cdots,y_{N_\text{f}}-\Delta^{(\beta)}_{N_\text{f}}}(N-1-N')
\right]
\label{Nfgeneralevenrecursionrelation}
\end{align}
for even $N_\text{f}$, and
\begin{align}
&Z_{y_1,\cdots,y_{N_\text{f}}}^{(N_\text{f})}(N)=\frac{1}{2\cosh\frac{(\zeta+m)N}{2}-(-1)^N2\cosh\frac{(\zeta-m)N}{2}}\nonumber \\
&\quad \times \left[ -\sum_{N'=1}^{N-1}\left(e^{\frac{(\zeta+m)(2N'-N)}{2}}
-(-1)^N e^{-\frac{(\zeta-m)(2N'-N)}{2}}
\right)Z_{y_1,\cdots,y_{N_\text{f}}}^{(N_\text{f})}(N')Z_{y_1,\cdots,y_{N_\text{f}}}^{(N_\text{f})}(N-N')\right.\nonumber \\
&\quad \quad \quad +
i^{N_\text{f}-1}\sum_{\beta=1}^{N_\text{f}}\frac{R_{y_\beta}^{-1}}{\prod_{\gamma(\neq\beta)}2\sinh\frac{y_\beta-y_\gamma}{2}}\sum_{N'=0}^{N-1}\left(-iR_m^{\frac{1}{N_\text{f}}}e^{\frac{1}{2}(1-\frac{1}{N_\text{f}})\zeta}\right)^{2N'-N+1}\nonumber \\
&\quad\quad\quad\times \left.Z^{(N_{\text{f}})}_{y_1+\Delta^{(\beta)}_1,\cdots,y_{N_\text{f}}+\Delta^{(\beta)}_{N_\text{f}}}(N')
Z^{(N_{\text{f}})}_{y_1-\Delta^{(\beta)}_1,\cdots,y_{N_\text{f}}-\Delta^{(\beta)}_{N_\text{f}}}(N-1-N')
\right]
\label{Nfgeneraloddrecursionrelation}
\end{align}
for odd $N_\text{f}$.
We find that the partition functions at $N=3$ generated by these recursion relations are consistent with the actual exact values calculated by the residue sum \eqref{totalZincombinatorialformula} with \eqref{ZNfcombinatorial2}.
In addition, the partition functions at finite but large $N$ generated by the recursion relations show a good agreement with the Airy form \eqref{ZNfpertAiry2}, as displayed in table \ref{recursionvsAiry}.
\begin{table}
\begin{center}
\begin{align*}
\begin{tabular}{|c|c|c|c|}
\hline
$N$&$\frac{Z^{(2)}_y(N)}{Z^{(2)}_{\text{pert},y_\alpha}(N)}$&$\frac{Z^{(3)}_{y_\alpha}(N)}{Z^{(3)}_{\text{pert},y_\alpha}(N)}$&$\frac{Z^{(4)}_{y_\alpha}(N)}{Z^{(4)}_{\text{pert},y_\alpha}(N)}$\\ \hline
$5$&$0.9981889472806604$&$0.9987219607832177$&$0.9980433319631781$\\ \hline
$10$&$0.9998020391863240$&$0.9998867022562782$&$0.9998146465395487$\\ \hline
$15$&$0.9999653482367743$&$0.9999832326328275$&$0.9999712381636826$\\ \hline
$20$&$0.9999921443116418$&$0.9999967098115862$&-\\ \hline
$30$&$0.9999993591068216$&$0.9999997902357814$&-\\ \hline
$40$&$0.9999999231954595$&-                   &-\\ \hline
\end{tabular}
\end{align*}
\caption{
The comparison between the partition function $Z^{(N_\text{f})}_{y_\alpha}(N)$ of super Yang-Mills theory with $N_\text{f}=2,3,4$ flavors generated by the recursion relation \eqref{Nf2recursionrelation} and the Airy form $Z^{(N_\text{f})}_{\text{pert},y_\alpha}(N)$ \eqref{ZNfpertAiry2} with \eqref{CBA}.
The parameters $\zeta,m$ are chosen as $\zeta=4i$, $m=\frac{i}{2}$.
The parameters $y_\alpha$ are chosen as $y_1=-y_2=\frac{1}{1000}$ for $N_\text{f}=2$, $(y_1,y_2,y_3)=(\frac{1}{1000},\frac{1}{500},-\frac{3}{1000})$ for $N_\text{f}=3$ and $(y_1,y_2,y_3,y_4)=(
\frac{1}{1000},\frac{1}{500},\frac{3}{1000},-\frac{3}{500})$ for $N_\text{f}=4$.
The results for $N_\text{f}=3$ at $N=40$ and $N_\text{f}=4$ at $N\ge 20$ are omitted due to the lack of our computational resource.
}
\label{recursionvsAiry}
\end{center}
\end{table}
These results strongly suggest that the bilinear equation \eqref{generalevenNf} and \eqref{generaloddNf} holds at all order in $u$.

\subsection{Evidence for general combinatorial formula \eqref{ZNfcombinatorial2}}
\label{sec_evidenceforcombinatorialformula}

Once we assume the bilinear equations \eqref{generalevenNf} and \eqref{generaloddNf}, we can also test the combinatorial formula \eqref{totalZincombinatorialformula} with \eqref{ZNfcombinatorial2} for higher ranks $N\ge 4$, which contain the Young diagrams of non-hook type.
For concreteness, here we demonstrate this analysis only for $N_\text{f}=2$, although the same analysis can be done for general $N_\text{f}$.

For $N_\text{f}=2$ and $N=4$, by using the formula \eqref{powerofRmandRy} we can list the powers of $R_m$ and $R_y$ appearing in the partition function and the corresponding pairs of Young diagram $(\lambda^{(1)},\lambda^{(2)})$ as
\begin{align}
&R_m^{-6}R_y^{-4}\rightarrow \lambda^{(\alpha)}=(\ydiagram{4},\emptyset),\quad
R_m^{-6}R_y^{4}\rightarrow \lambda^{(\alpha)}=(\emptyset,\ydiagram{4}),\quad
R_m^{-3}R_y^{-2}\rightarrow \lambda^{(\alpha)}=(\ydiagram{3},\ydiagram{1}),\nonumber \\
&R_m^{-3}R_y^{2}\rightarrow \lambda^{(\alpha)}=(\ydiagram{1},\ydiagram{3}),\quad
R_m^{-2}R_y^{-4}\rightarrow \lambda^{(\alpha)}=(\ydiagram{3,1},\emptyset),\quad
R_m^{-2}R_y^{0}\rightarrow \lambda^{(\alpha)}=(\ydiagram{2},\ydiagram{2}),\nonumber \\
&R_m^{-2}R_y^{4}\rightarrow \lambda^{(\alpha)}=(\emptyset,\ydiagram{3,1}),\quad
R_m^{0}R_y^{-4}\rightarrow \lambda^{(\alpha)}=(\ydiagram{2,2},\emptyset),\quad
R_m^{0}R_y^{-2}\rightarrow \lambda^{(\alpha)}=(\ydiagram{2,1},\ydiagram{1}),\nonumber \\
&R_m^{0}R_y^{0}\rightarrow \lambda^{(\alpha)}=(\ydiagram{2},\ydiagram{1,1}),(\ydiagram{1,1},\ydiagram{2}),\quad
R_m^{0}R_y^{2}\rightarrow \lambda^{(\alpha)}=(\ydiagram{1},\ydiagram{2,1}),\quad
R_m^{0}R_y^{4}\rightarrow \lambda^{(\alpha)}=(\emptyset,\ydiagram{2,2}),\nonumber \\
&R_m^{2}R_y^{-4}\rightarrow \lambda^{(\alpha)}=(\ydiagram{2,1,1},\emptyset),\quad
R_m^{2}R_y^{0}\rightarrow \lambda^{(\alpha)}=(\ydiagram{1,1},\ydiagram{1,1}),\quad
R_m^{2}R_y^{4}\rightarrow \lambda^{(\alpha)}=(\emptyset,\ydiagram{2,1,1}),\nonumber \\
&R_m^{3}R_y^{-2}\rightarrow \lambda^{(\alpha)}=(\ydiagram{1,1,1},\ydiagram{1}),\quad
R_m^{3}R_y^{2}\rightarrow \lambda^{(\alpha)}=(\ydiagram{1},\ydiagram{1,1,1}),\quad
R_m^{6}R_y^{-4}\rightarrow \lambda^{(\alpha)}=(\ydiagram{1,1,1,1},\emptyset),\nonumber \\
&R_m^{6}R_y^{4}\rightarrow \lambda^{(\alpha)}=(\emptyset,\ydiagram{1,1,1,1}).
\end{align}
By calculating $Z^{(2)}_y(4)$ by using the recursion relation \eqref{Nfgeneralevenrecursionrelation}, which reduces for $N_\text{f}=2$ to
\begin{align}
&Z_y^{(2)}(N)=\frac{1}{2\cosh\frac{(\zeta+m+\pi i)N}{2}-2\cosh\frac{(\zeta-m-\pi i)N}{2}}\nonumber \\
&\quad \times \left[ -\sum_{N'=1}^{N-1}(e^{\frac{(\zeta+m+\pi i)(2N'-N)}{2}}
-e^{-\frac{(\zeta-m-\pi i)(2N'-N)}{2}}
)Z_y^{(2)}(N')Z_y^{(2)}(N-N')\right.\nonumber \\
&\quad\quad\ +\frac{1}{2\sinh y}\sum_{N'=0}^{N-1}
\left(
R_y(R_m^{\frac{1}{2}}e^{\frac{\zeta}{4}})^{2N'-N+1}
-R_y^{-1}(R_m^{\frac{1}{2}}e^{\frac{\zeta}{4}})^{-2N'+N-1}
\right)\nonumber \\
&\quad\quad\quad\quad\left.\times Z^{(2)}_{y+\frac{m+\pi i}{2}}(N')
Z^{(2)}_{y-\frac{m+\pi i}{2}}(N-1-N')
\right],
\label{Nf2recursionrelation}
\end{align}
we can read off the building blocks with non-hook Young diagram$Z^{(2)}_{\ydiagram{2,2},\emptyset}$ and $Z^{(2)}_{\emptyset,\ydiagram{2,2}}$ from the coefficients of $R_m^0R_y^{\mp 4}$.
As a result, we find that these building blocks are precisely given by the combinatorial formula \eqref{ZNfcombinatorial2}.
For $N=5$, the powers of $R_m$ and $R_y$ to which non-hook type Young diagrams contribute are
\begin{align}
&R_m^{-2}  R_y^{-5 }\rightarrow  \lambda^{(\alpha)}=( \ydiagram{3,2},\emptyset),\quad
R_m^{-2}  R_y^{5  }\rightarrow  \lambda^{(\alpha)}=(\emptyset,\ydiagram{3,2}),\quad
R_m^{0 }  R_y^{-3 }\rightarrow  \lambda^{(\alpha)}=( \ydiagram{2,2},\ydiagram{1}),\nonumber \\
&R_m^{0 }  R_y^{3  }\rightarrow  \lambda^{(\alpha)}=(\ydiagram{1},\ydiagram{2,2}),\quad
R_m^{2 }  R_y^{-5 }\rightarrow  \lambda^{(\alpha)}=( \ydiagram{2,2,1},\emptyset),\quad
R_m^{2 }  R_y^{5  }\rightarrow  \lambda^{(\alpha)}=(\emptyset,\ydiagram{2,2,1}).
\end{align}
Again, we find that the building blocks with non-hook Young diagrams read off from the coefficients of $Z^{(2)}_y(5)$ obtained by the recursion relation are consistent with the combinatorial formula \eqref{ZNfcombinatorial2}.
For $N=6$, the powers of $R_m$ and $R_y$ to which non-hook type Young diagrams contribute are
\begin{align}
&R_m^{-5}R_y^{-6}\rightarrow\lambda^{(\alpha)}=(\ydiagram{4,2},\emptyset),\quad
R_m^{-5}R_y^{6}\rightarrow\lambda^{(\alpha)}=(\emptyset,\ydiagram{4,2}),\nonumber \\
&R_m^{-3}R_y^{-6}\rightarrow\lambda^{(\alpha)}=(\ydiagram{4,1,1},\emptyset),(\ydiagram{3,3},\emptyset),\quad
R_m^{-3}R_y^{6}\rightarrow\lambda^{(\alpha)}=(\emptyset,\ydiagram{4,1,1}),(\emptyset,\ydiagram{3,3}),\nonumber \\
&R_m^{-2}R_y^{-4}\rightarrow\lambda^{(\alpha)}=(\ydiagram{3,2},\ydiagram{1}),\quad
R_m^{-2}R_y^{4}\rightarrow\lambda^{(\alpha)}=(\ydiagram{1},\ydiagram{3,2}),\quad
R_m^{-1}R_y^{-2}\rightarrow\lambda^{(\alpha)}=(\ydiagram{3,1},\ydiagram{1,1}),(\ydiagram{2,2},\ydiagram{2}),\nonumber \\
&R_m^{-1}R_y^{2}\rightarrow\lambda^{(\alpha)}=(\ydiagram{2},\ydiagram{2,2}),(\ydiagram{1,1},\ydiagram{3,1}),\quad
R_m^{0}R_y^{-6}\rightarrow\lambda^{(\alpha)}=(\ydiagram{3,2,1},\emptyset),\quad
R_m^{0}R_y^{6}\rightarrow\lambda^{(\alpha)}=(\emptyset,\ydiagram{3,2,1}),\nonumber \\
&R_m^{1}R_y^{-2}\rightarrow\lambda^{(\alpha)}=(\ydiagram{2,2},\ydiagram{1,1}),(\ydiagram{2,1,1},\ydiagram{2}),\quad
R_m^{1}R_y^{2}\rightarrow\lambda^{(\alpha)}=(\ydiagram{2},\ydiagram{2,1,1}),(\ydiagram{1,1},\ydiagram{2,2}),\nonumber \\
&R_m^{2}R_y^{-4}\rightarrow\lambda^{(\alpha)}=(\ydiagram{2,2,1},\ydiagram{1}),\quad
R_m^{2}R_y^{4}\rightarrow\lambda^{(\alpha)}=(\ydiagram{1},\ydiagram{2,2,1}),\quad
R_m^{3}R_y^{-6}\rightarrow\lambda^{(\alpha)}=(\ydiagram{3,1,1,1},\emptyset),(\ydiagram{2,2,2},\emptyset),\nonumber \\
&R_m^{3}R_y^{6}\rightarrow\lambda^{(\alpha)}=(\emptyset,\ydiagram{3,1,1,1}),(\emptyset,\ydiagram{2,2,2}),\quad
R_m^{5}R_y^{-6}\rightarrow\lambda^{(\alpha)}=(\ydiagram{2,2,1,1},\emptyset),\quad
R_m^{5}R_y^{6}\rightarrow\lambda^{(\alpha)}=(\emptyset,\ydiagram{2,2,1,1}).
\end{align}
When the power is in one-to-one correspondence with a pair of Young diagram, we can follow the same strategy as in $N=4,5$ and find agreement with the combinatorial formula \eqref{ZNfcombinatorial2}.
Furthermore, even when two pairs of Young diagram contribute in the same powers of $R_m$ and $R_y$, since one of them is of hook-type, whose building can be calculated by \eqref{ZNfcombinatorial2} without any subtlety, we can still determine the building blocks for non-hook Young diagrams from $Z^{(2)}_y(6)$ obtained by the recursion relation.
These buidling blocks also turn out to be consistent with the combinatorial formula \eqref{ZNfcombinatorial2}.
We can continue this analysis until two or more pairs of Youg diagrams containing of non-hook type contribute in the same powers of $R_m$ and $R_y$, which strongly support the validity of the combinatorial formula \eqref{ZNfcombinatorial2} for the non-hook type Young diagrams.

\section{Application: leading non-perturbative effect in $1/N$}
\label{sec_1stwscoef}

In the previous section we have obtained new recursion relations \eqref{Nfgeneralevenrecursionrelation} and \eqref{Nfgeneraloddrecursionrelation} with respect to $N$ for the $N_\text{f}$ matrix model \eqref{ADHMgeneralNf}.
As an application of these recursion relations, in this section we study the leading non-perturbative correction in $1/N$ to the free energy $-\log Z^{(N_\text{f})}_{y_\alpha}(N)$ of the $N_\text{f}$ matrix model.

The non-perturbative effects in $N_\text{f}$ matrix model were also studied in \cite{Nosaka:2015iiw} for general $N_\text{f},\zeta,m$ with $y_\alpha=0$ through the ${\cal N}=4$ $\text{U}(N)_k\times \text{U}(N)_0^{\otimes (N_\text{f}-1)}\times \text{U}(N)_{-k}$ circular quiver Chern-Simons theory which is $\text{SL}(2,\mathbb{Z})$ dual to the super Yang-Mills theory with $N_\text{f}$ hypermultiplets when $k=1$.
In \cite{Nosaka:2015iiw} the non-perturbative effects are given in terms of the modified grand potential $J(\mu)$ defined by
\begin{align}
\sum_{n=0}^\infty e^{\mu N}Z_{y_\alpha}^{(N_\text{f})}(N)=\sum_{n=-\infty}^\infty e^{J(\mu+2\pi in)},
\end{align}
as
\begin{align}
J(\mu)=J_\text{pert}(\mu)+J_\text{np}(\mu)
\end{align}
with
\begin{align}
J_\text{pert}(\mu)=\frac{C}{3}\mu^3+B\mu+A,\quad
J_\text{np}(\mu)=\sum_{\omega}d_\omega e^{-\omega \mu}.
\label{JpertJnp}
\end{align}
Here $C$, $B$ and $A$ are given in \eqref{CBA} and $d_\omega$ are some functions of $\zeta,m$ which are independent of $\mu$, and the possible values of $\omega$ are the linear combinations of the following exponents with non-negative integer coefficients
\begin{align}
\omega_{\text{MB}1\pm},\quad
\omega_{\text{MB}2\pm},\quad
\omega_{\text{MB}3},\quad
\omega_{\text{WS}\pm\pm'},
\end{align}
where $\omega_{\text{MB}1\pm}$, $\omega_{\text{MB}2\pm}$ and $\omega_{\text{MB}3}$ are given as
\begin{align}
\omega_{\text{MB}1\pm}=\frac{2}{N_\text{f}\pm\frac{i\zeta}{\pi}},\quad
\omega_{\text{MB}2\pm}=\frac{2}{1\pm\frac{im}{\pi}},\quad
\omega_{\text{MB}3}=1,
\label{omegaMB}
\end{align}
while $\omega_{\text{WS}\pm\pm'}$ were proposed as
\begin{align}
\omega_{\text{WS}\pm\pm'}=\frac{4}{k(N_\text{f}\pm\frac{i\zeta}{\pi})(1\pm'\frac{im}{\pi})},\quad k=1.
\label{omegaWSpmpm}
\end{align}
Indeed, the modified grand potential and the original partition function are related through the inverse transformation
\begin{align}
Z^{(N_\text{f})}_{y_\alpha}(N)=\int_{-i\infty}^{i\infty}\frac{d\mu}{2\pi i}e^{J(\mu)-\mu N}.
\end{align}
and substituting \eqref{JpertJnp} to this formula and expanding $e^{J_\text{np}(\mu)}$ as
\begin{align}
e^{J_\text{np}(\mu)}=
\sum_{n=0}^\infty\frac{1}{n!}\Bigl(\sum_{\omega}d_\omega e^{-\omega\mu}\Bigr)^n,
\end{align}
we obtain
\begin{align}
Z(N)=
e^AC^{-\frac{1}{3}}\text{Ai}[C^{-\frac{1}{3}}(N-B)]
+\sum_\omega d_\omega e^AC^{-\frac{1}{3}}\text{Ai}[C^{-\frac{1}{3}}(N+\omega-B)]
+\cdots.
\label{Zfrominversetrsfwithnp}
\end{align}
Taking into account of the asymptotics of the Airy function $\text{Ai}(z)$ at $z\rightarrow\infty$, $\text{Ai}(z)\sim e^{-\frac{2}{3}z^{\frac{3}{2}}}$, we find that the second term in \eqref{Zfrominversetrsfwithnp} gives the following non-perturbative correction in the large $N$ expansion of the free energy $-\log Z^{(N_\text{f})}_{y_\alpha}(N)$
\begin{align}
-\log Z^{(N_\text{f})}_{y_\alpha}(N)=-\log Z^{(N_\text{f})}_{\text{pert},y_\alpha}(N)+\sum_\omega d_\omega e^{-\omega \sqrt{\frac{N-B}{C}}}+\cdots.
\end{align}

These non-perturbative effects can be interpreted in the conformal limit $\zeta=m=y_\alpha=0$ as the contributions of closed M2-branes wrapped on three-dimensional subspaces of the internal space $Y_7$ of the dual geometry $\text{AdS}_4\times Y_7$.
In particular, from the viewpoint of the superconformal Chern-Simons theory with general $k$, where $Y_7$ contains a circular direction modded by $\mathbb{Z}_k$, the exponents $\omega_{\text{WS}\pm\pm'}$ (resp.~$\omega_{\text{MB}1\pm}$, $\omega_{\text{MB}2\pm}$ and $\omega_{\text{MB}3}$) are associated with the M2-branes whose worldvolumes do (resp.~do not) contain the $\mathbb{Z}_k$-modded circle.
Since the $\mathbb{Z}_k$-modded circle can also be regarded as the M-theory direction, we shall refer the non-perturbative effects with the exponent $\omega_{\text{WS}\pm\pm'}$ (resp.~$\omega_{\text{MB}1\pm}$, $\omega_{\text{MB}2\pm}$ and $\omega_{\text{MB}3}$) to ``worldsheet instantons'' (resp.~membrane instantons).
In this section, we analyze the non-perturbative effects by fitting the deviation of the exact values, which can be generated by solving the recursion relations \eqref{Nfgeneralevenrecursionrelation} and \eqref{Nfgeneraloddrecursionrelation} numerically, from the perturbative part $Z^{(N_\text{f})}_{\text{pert},y_\alpha}(N)$ \eqref{ZNfpertAiry2}.
In particular, we confirm the proposed exponents of the worldsheet instantons \eqref{omegaWSpmpm} and also determine their instanton coefficients $d_{\omega_{\text{WS}\pm\pm'}}$ as a function of $\zeta$, $m$ and $y_\alpha$ as
\begin{align}
d_{\omega_{\text{WS}\pm\pm'}}^{(N_\text{f})}(\zeta,m;y_\alpha)=\frac{1}{4
\sin\frac{2\pi}{N_\text{f}\pm\frac{i\zeta}{\pi}}
\sin\frac{2\pi}{1\pm'\frac{im}{\pi}}
}
\sum_{\alpha=1}^{N_\text{f}}e^{\pm\frac{2y_\alpha}{1\pm'\frac{im}{\pi}}}.
\label{dNfpmpmconjecture}
\end{align}

\subsection{Known results}

Let us first review the known results in the related setups.
These results play crucial role in determining $\zeta,m,y_\alpha$-dependence of the leading non-perturbative effect in the $N_\text{f}$ matrix model.

First the $\text{U}(N)_k\times \text{U}_0^{\otimes (N_\text{f}-1)}\times \text{U}(N)_{-k}$ circular quiver Chern-Simons theories was also studied in \cite{Hatsuda:2015lpa} without mass and FI deformation (which corresponds to $\zeta=m=y_\alpha=0$ in the $N_\text{f}$ matrix model), where the leading worldsheet instanton effect was found as
\begin{align}
J_{\text{np}}=
\frac{N_\text{f}}{\sin\frac{2\pi}{N_\text{f}k}\sin\frac{2\pi}{k}}
e^{-\frac{4}{N_\text{f}k}\mu}
+\cdots.
\label{generalNfwithzeta0m0y0}
\end{align}
Since all the four worldsheet instanton exponents \eqref{omegaWSpmpm} coincides with $\frac{4}{N_\text{f}k}$ in the limit $\zeta,m\rightarrow 0$, this result of \cite{Hatsuda:2015lpa} should be understood as net contribution of the four worldsheet instantons.

Second, when $N_\text{f}=1$, the $N_\text{f}$ matrix model is equivalent to the $\text{U}(N)_k\times \text{U}(N+M)_{-k}$ ABJM theory for $k=1$ and $M=0$ with two-parameter real mass deformation corresponding to $\zeta$ and $m$.
The M2-instanton corrections in this model was studied in \cite{Nosaka:2024gle}, where it was found that coefficients of the leading worldsheet instantons are given as
\begin{align}
J_{\text{np}}=\sum_{\pm,\pm'}\frac{\cos\frac{2\pi M}{k}}{4
\sin\frac{2\pi}{k(1\pm\frac{i\zeta}{\pi})}
\sin\frac{2\pi}{k(1\pm'\frac{im}{\pi})}
}e^{-\frac{4}{k(1\pm\frac{i\zeta}{\pi})(1\pm'\frac{im}{\pi})}\mu}
+\cdots.
\label{Nf1withy0}
\end{align}
Note that these four coefficients can be viewed as the resolution of the first worldsheet instanton coefficient of the ABJ theory without mass deformation \cite{Hatsuda:2012dt,Matsumoto:2013nya}
\begin{align}
J_{\text{np}}=\frac{\cos\frac{2\pi M}{k}}{\sin^2\frac{2\pi}{k}}e^{-\frac{4\mu}{k}}+\cdots.
\label{d1inABJ}
\end{align}

Note that the $M$-dependence of the instanton coefficient of the ABJ theory \eqref{d1inABJ} (and its mass deformation \eqref{Nf1withy0}) appears only in the numerator as a Laurent polynomial of $e^{\frac{2\pi iM}{k}}$ for the following reason.
Under the TS/ST correspondence, the worldsheet instanton effects in the ABJ theory are given by the Gopakumar-Vafa free energy \cite{Gopakumar:1998jq} of the topological string theory on local $\mathbb{P}^1\times \mathbb{P}^1$
\begin{align}
F_\text{top}(T_1,T_2)=\sum_{g\ge 0}\sum_{w\ge 1}\sum_{\substack{d_1,d_2\ge 0\\ ((d_1,d_2)\neq (0,0))}}\frac{1}{w}n^{d_1,d_2}_g\Bigl(2\sin\frac{2\pi}{k}\Bigr)^{2g-2}e^{-w(d_1T_1+d_2T_2)}
\end{align}
where $n^{d_1,d_2}_g$ are the Gopakumar-Vafa invariants and \cite{Honda:2014npa}\footnote{
For simplicity, here we ignore the difference between $\mu$ and the effective chemical potential $\mu_\text{eff}$ which takes care of the bound states of the worldsheet instantons and membrane instantons.
}
\begin{align}
T_1=\frac{4\mu}{k}+\frac{2\pi i}{k}\Bigl(\frac{k}{2}-M\Bigr),\quad
T_2=\frac{4\mu}{k}-\frac{2\pi i}{k}\Bigl(\frac{k}{2}-M\Bigr).
\end{align}
Hence the $M$-dependent factor in the instanton coefficient \eqref{d1inABJ} comes from the devaitions of the Kahler parameters $T_I$ from $\frac{4\mu}{k}$.
Note that the Kahler parameters are related to the coefficients of the inverse of the density matrix \cite{Bonelli:2017gdk}
\begin{align}
\hat{\rho}^{-1}=e^{{\hat x}}+e^{{\hat p}}+e^{-{\hat p}}+e^{\pi i(k-2M)}e^{-{\hat x}},\quad [{\hat x},{\hat p}]=i\hbar,\quad (\hbar=\pi k),
\end{align}
which gives the mirror curve of the Calabi-Yau threefold, as\footnote{
The rescaling by $\frac{2\pi}{\hbar}$ was pointed out in e.g.~\cite{Kashaev:2015wia,Bonelli:2017gdk}.
}
$T_I\sim \frac{2\pi}{\hbar}\log (\text{coefficients})$.

\subsection{Leading instantons in $N_\text{f}$ matrix models}

\begin{figure}
\begin{center}
\includegraphics[width=\textwidth]{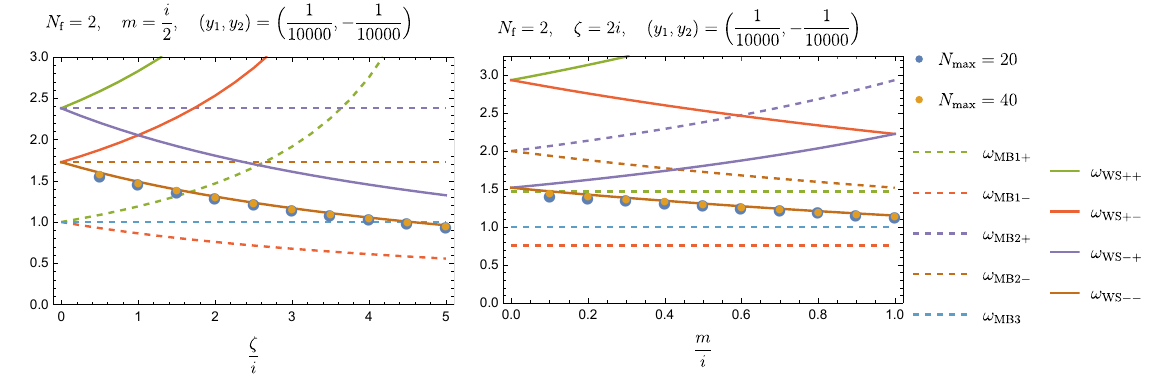}
\includegraphics[width=\textwidth]{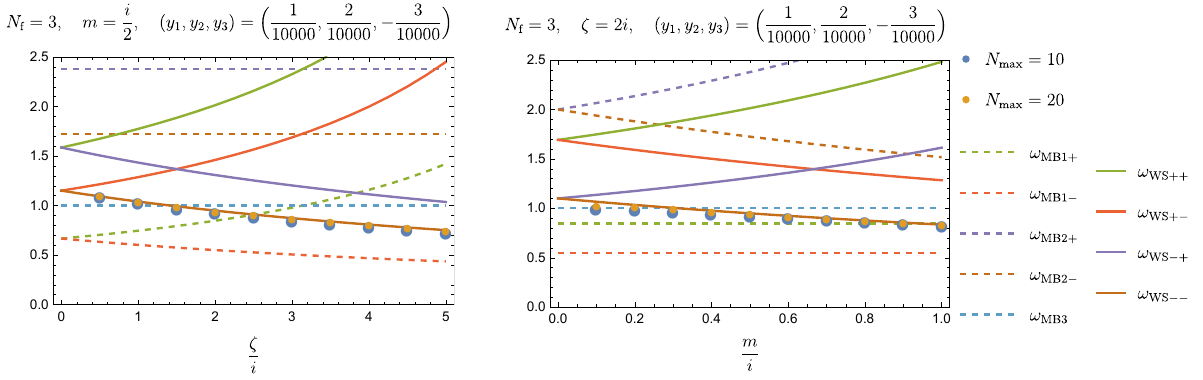}
\caption{
Comparison between the leading instanton exponent $\omega$ obtained by fittng with ansatn \eqref{instexpfitting} and the membrane/worldsheet instanton exponents \eqref{omegaMB} and \eqref{omegaWSpmpm} proposed in \cite{Nosaka:2015iiw}.
}
\label{1stinstexpNf2and3}
\end{center}
\end{figure}

Now we investigate the leading M2-instanton effects in $N_\text{f}$ matrix models.
First let us consider the cases with $y_\alpha=0$.
We can study the exponent of the leading M2-instanton exponent by fitting the exact values of the partition function with an ansatz
\begin{align}
\frac{Z^{(N_\text{f})}_{y_\alpha}(N)}{Z^{(N_\text{f})}_{\text{pert},y_\alpha}(N)}-1\approx d e^{-\omega \sqrt{\frac{N-B}{C}}},
\label{instexpfitting}
\end{align}
with the fitting parameters $d$ and $\omega$.
In figure \ref{1stinstexpNf2and3} we compare the values of $\omega$ obtained by the fitting for $\zeta,m\in i\mathbb{R}_{>0}$ with the membrane instanton exponents \eqref{omegaMB} and the worldsheet instanton exponents \eqref{omegaWSpmpm} proposed in \cite{Nosaka:2015iiw} with $k=1$.
Interestingly, the results of fitting show a good agreement with the worldsheet instanton exponent $\omega_{\text{WS}--}$ which is always larger than the membrane instanton exponent $\omega_{\text{MB}1-}$ and is also larger than the membrane instanton exponents $\omega_{\text{MB}1+}$ and $\omega_{\text{MB}3}$ depending on the values of $\zeta$ and $m$.
This implies that for $k=1$, which corresponds to the $N_\text{f}$ matrix model, the coefficients of the membrane instantons with the exponents $\omega_{\text{MB}1\pm}$ or $\omega_{\text{MB}3}$ vanishes and hence the actual leading non-perturbative effect is given by the worldsheet instanton with the exponent $\omega_{\text{WS}--}$.

Taking into account the expression of the worldsheet instanton coefficient for $N_\text{f}=1$ \eqref{Nf1withy0} and the expression for general $N_\text{f}$ with $\zeta=m=0$ \eqref{generalNfwithzeta0m0y0} (which should split into four coefficients once $\zeta,m$ are turned on), we guess that the worldsheet instanton coefficient $d^{(N_\text{f})}_{\omega_{\text{WS}--}}$ for $y_\alpha=0$ is given as
\begin{align}
d_{\omega_{\text{WS}--}}^{(N_\text{f})}(\zeta,m;0)
=\frac{N_\text{f}}{4\sin\frac{2\pi}{N_\text{f}-\frac{i\zeta}{\pi}}\sin\frac{2\pi}{1-\frac{im}{\pi}}}.
\end{align}
We have confirmed this formula for $N_\text{f}=2$ and $N_\text{f}=3$ through the fitting of the finite $N$ exact values of the partition functions.

Now let us turn on the mass parameters $y_\alpha$ of the fundamental hypermultiplets.
In order to make a guess of the $y_\alpha$ dependence of the non-perturbative effects, we notice that the $N_\text{f}$ matrix model \eqref{ADHMgeneralNf} can also be written in the Fermi gas formalism, whose density matrix is
\begin{align}
\hat{\rho}^{(N_\text{f})}_{y_\alpha}=\frac{e^{-\frac{i\zeta}{2\pi}{\hat x}}}{\prod_{\alpha=1}^{N_\text{f}}2\cosh\frac{{\hat x}-y_\alpha}{2}}\frac{e^{\frac{im}{2\pi}{\hat p}}}{2\cosh\frac{{\hat p}}{2}},\quad
([{\hat x},{\hat p}]=2\pi i).
\end{align}
In particular, $y_\alpha$ enter in the inverse of the density matrix only through the coefficients of the Laurent polynomial of $e^{\frac{{\hat x}}{2}}$, $e^{\frac{i\zeta}{2\pi}{\hat x}}$, $e^{\frac{{\hat p}}{2}}$ and $e^{-\frac{im}{2\pi}{\hat p}}$ in the form $e^{\pm\frac{y_\alpha}{2}}$.
Therefore, taking into account of the TS/ST correspondence,\footnote{
Note that for generic values of $\zeta$ and $m$, the inverse density matrix $({\hat\rho}^{(N_\text{f})}_{y_\alpha})^{-1}$ may not be associated with any toric Calabi-Yau threefold.
When $\zeta,m\in\pi i\mathbb{Q}$, however, after an appropriate rescaling of the canonical position/momentum operators $({\hat x},{\hat p})$ into $({\hat x}_\text{new},{\hat p}_\text{new})$ we can express $({\hat\rho}^{(N_\text{f})}_{y_\alpha})^{-1}$ as a finite Laurent polynomial of $e^{{\hat x}_{\text{new}}}$ and $e^{{\hat p}_{\text{new}}}$, and hence we expect that the prescription of the TS/ST correspondence for a higher genus mirror curve \cite{Codesido:2015dia} is applicable (see e.g.~\cite{Nosaka:2024gle}).
In this viewpoint, the $m$-dependence of the numerator in the conjectured worldsheet instanton coefficient $d_{\pm\pm'}^{(N_\text{f})}(\zeta,m;y_\alpha)$ \eqref{dNfpmpmconjecture} might be related to the fact that the Planck constant $\hbar_\text{new}$ defined by the canonical commutation relation of the rescaled operators $[{\hat x}_\text{new},{\hat p}_\text{new}]=i\hbar_\text{new}$ depends non-trivially on $\zeta$ and $m$.
}
we expect that the $y_\alpha$-dependence of the worldsheet instanton coefficient is of the similar form as the $M$-dependence of the worldsheet instanton coefficient of the $\text{U}(N)_k\times \text{U}(N+M)_{-k}$ ABJ theory which we mentioned below \eqref{d1inABJ}.
More concretely, we guess
\begin{align}
d_{\omega_{\text{WS}--}}^{(N_\text{f})}(\zeta,m;y_\alpha)
=\frac{
\sum_{\alpha=1}^{N_\text{f}} e^{f(\zeta,m)y_\alpha}
}{4\sin\frac{2\pi}{N_\text{f}-\frac{i\zeta}{\pi}}\sin\frac{2\pi}{1-\frac{im}{\pi}}},
\end{align}
with $f(\zeta,m)$ some simple rational functions of $\zeta$ and $m$.

\begin{figure}
\begin{center}
\includegraphics[width=0.45\textwidth]{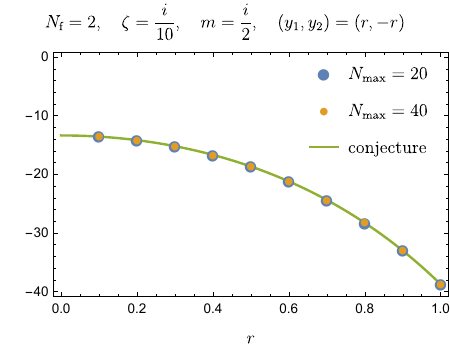}
\includegraphics[width=0.45\textwidth]{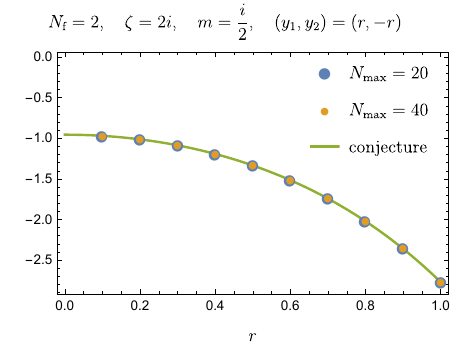}
\caption{
Comparison between the leading instanton coefficient $d_{\omega_{\text{WS}--}}^{(2)}$ for $N_\text{f}=2$ obtained by fitting and the conjectured expression \eqref{Nf2conjecture}.
}
\label{1stinstcoefNf2}
\end{center}
\end{figure}

\begin{figure}
\begin{center}
\includegraphics[width=0.45\textwidth]{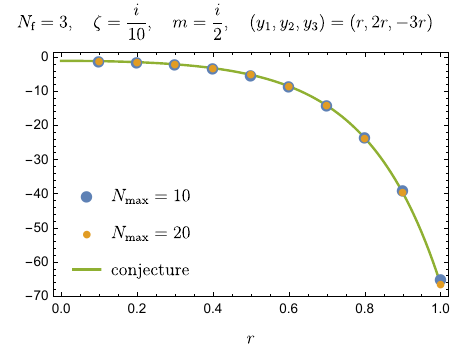}
\includegraphics[width=0.45\textwidth]{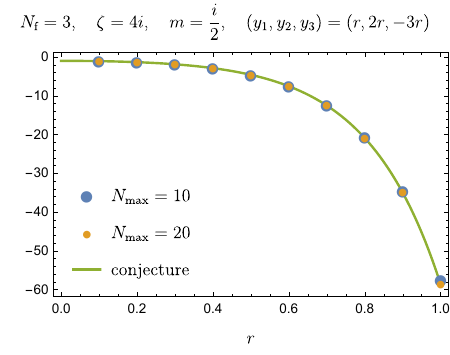}\\
\includegraphics[width=0.45\textwidth]{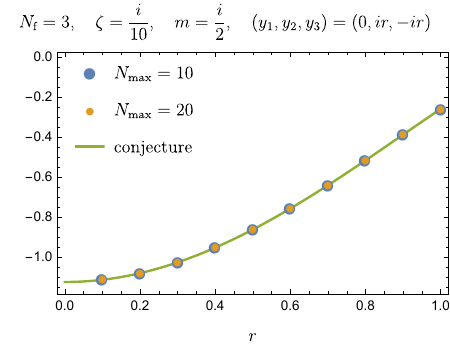}
\includegraphics[width=0.45\textwidth]{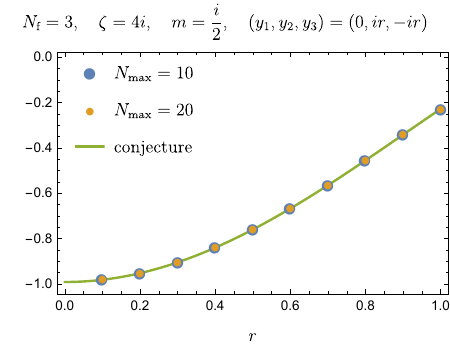}\\
\includegraphics[width=0.45\textwidth]{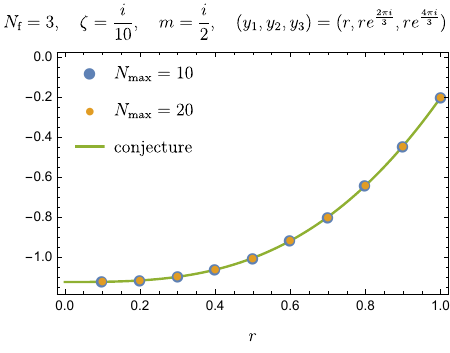}
\includegraphics[width=0.45\textwidth]{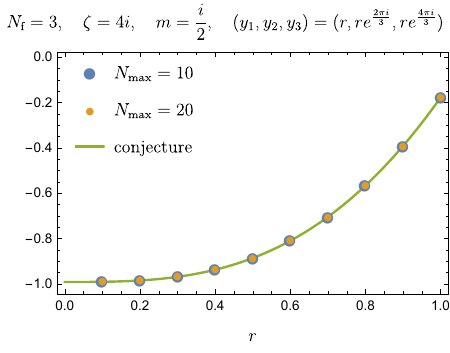}
\caption{
Comparison between the leading instanton coefficient $d_{\omega_{\text{WS}--}}^{(3)}$ for $N_\text{f}=3$ obtained by fitting and the conjectured expression \eqref{Nf3conjecture}.
}
\label{1stinstcoefNf3}
\end{center}
\end{figure}

For $N_\text{f}=2$ with $y_1=-y_2=y$, we conjecture
\begin{align}
d_{\omega_{\text{WS}--}}^{(2)}(\zeta,m;y)=\frac{\cosh\frac{2y}{1-\frac{im}{\pi}}}{2\sin\frac{2\pi}{2-\frac{i\zeta}{\pi}}\sin\frac{2\pi}{1-\frac{im}{\pi}}}
\label{Nf2conjecture}
\end{align}
from the results of the fitting.
See figure \ref{1stinstcoefNf2}.
For $N_\text{f}=3$, we conjecture
\begin{align}
d_{\omega_{\text{WS}--}}^{(3)}(\zeta,m;y_\alpha)=\frac{1}{4\sin\frac{2\pi}{3-\frac{i\zeta}{\pi}}\sin\frac{2\pi}{1-\frac{im}{\pi}}}
(
e^{-\frac{2y_1}{1-\frac{im}{\pi}}}
+e^{-\frac{2y_2}{1-\frac{im}{\pi}}}
+e^{-\frac{2y_3}{1-\frac{im}{\pi}}}),
\label{Nf3conjecture}
\end{align}
where $y_3=-y_1-y_2$.
See figure \ref{1stinstcoefNf3}.

From the results for $N_\text{f}=2$ \eqref{Nf2conjecture} and $N_\text{f}=3$ \eqref{Nf3conjecture}, and also taking into account the invariance of the $N_\text{f}$ matrix model \eqref{ADHMgeneralNf} under the changes of the parameters $(\zeta,m;y_\alpha)\rightarrow (\zeta,-m;y_\alpha)$ and $(\zeta,m;y_\alpha)\rightarrow (-\zeta,m;-y_\alpha)$, we conjecture that the coefficient $d_{\omega_{\text{WS}\pm\pm'}}^{(N_\text{f})}$ of the worldsheet instanton with the exponent $\omega_{\text{WS}\pm\pm'}$ for general $N_\text{f}$ is given by \eqref{dNfpmpmconjecture}.

\section{Discussion}
\label{sec_Discuss}

In this paper we have considered the 3d ${\cal N}=4$ $\text{U}(N)$ Yang-Mills theory with one adjoint hypermultiplet and $N_\text{f}$ fundamental hypermultiplets, which describes a stack of $N$ M2-branes placed on $\mathbb{C}^2\times \mathbb{C}^2/\mathbb{Z}_{N_\text{f}}$.
In the large $N$ limit, this theory is dual to $\text{AdS}_4\times S^7/\mathbb{Z}_{N_\text{f}}$ under the AdS/CFT correspondence, where $S^7/\mathbb{Z}_{N_\text{f}}$ is a radial section of $\mathbb{C}^2\times \mathbb{C}^2/\mathbb{Z}_{N_\text{f}}$.
For the $S^3$ partition function of this theory deformed by the FI parameter $\zeta$, the mass parameter $m$ of the adjoint hypermultiplet and the mass parameters $y_\alpha$ of the fundamental hypermultiplets, we have proposed a bilinear relation satisfied by the grand canonical sum of the partition function, which can also be written in the form of the recursion relation for the partition function with respect to the rank $N$.
The recursion relation reproduces the finite $N$ exact values of the partition function calculated by the multi-variate residue calculus from the localization matrix model \eqref{ADHMgeneralNf}.
We have also confirmed that the numerical values of the partition function for finite but very large $N$ generated by the recursion relation are consistent with the Airy form \eqref{ZNfpertAiry2} which was obtained by the Fermi gas formalism of the matrix model.
The recursion relation can be a new powerful tool to study the large $N$ expansion of the partition function.
As an application we investigated the leading non-perturbative correction in $1/N$ and successfully determined the analytic expression of its exponent as well as the overall coefficient as functions of $\zeta,m,y_\alpha$.
When the mass and FI deformation parameters are small, these non-perturbative correction would be understood holographically as the contributions from the M2-branes wrapped on three-dimensional volumes in the internal space of a smoothly deformed $\text{AdS}_4\times S^7/\mathbb{Z}_{N_\text{f}}$.
We also find a simple combinatorial formula for the partition function, which is suggested through the residue sum calculation and have been confirmed through the recursion relation.

There are several possible future directions of the research related to the current results.
First, it would be an interesting future work to find the bilinear relation and the recursion relation in more general M2-brane matrix models.
In particular, although so far only for the matrix models which facilitates the Fermi gas formalism have been investigated, it is highly desirable to generalize these structures also to the M2-brane matrix models without Fermi gas formalism.
There are various arguments through such as the numerical values of the partition function for finite (but large) $N$ \cite{Bobev:2025ltz} or the explicit calculations of the first few sub-leading terms in the gravity side as quantum/higher derivative corrections \cite{Bhattacharyya:2012ye,Bobev:2023dwx,Bobev:2021oku} that the Airy form \eqref{ZNfpertAiry2} is a universal structure for a large class of the matrix models calculating $S^3$ (or squashed sphere $S^3_b$) partition function of the theories of M2-branes.
However, so far the Airy form was proved only for the models with the Fermi gas formalism.
The new recursion relation might shed new light on this problem.
For this purpose, one of the simplest model to investigate would be the squashed sphere partition function of the super Yang-Mills theory with the squashing parameter $b$ satisfying $b^2\in 2\mathbb{N}-1$ \cite{Hatsuda:2016uqa,Kubo:2024qhq}.
These model cannot be written in the Fermi gas form for a generic value of the mass parameter $m$ for the adjoint hypermultiplet.
However, the one-loop determinant of the hypermultiplet, which in general written in terms of the double sine functions, simplify to finite products hyperbolic functions for $b^2\in 2\mathbb{N}-1$, and hence the residue sum calculation would be relatively tractable \cite{Kubo:2024qhq}.

Although in this paper we have analyzed the large $N$ behavior of the $N_\text{f}$ matrix models by solving the recursion relations numerically, it is more desirable to understand how to extract the large $N$ expansion analytically from the recursion relations.
In particular, it would be interesting if we can derive the Airy form of the large $N$ partition function \eqref{ZNfpertAiry2} directly from the recursion relations rather than through the Fermi gas formalism.

The bilinear relations for the M2-brane matrix models discovered so far passed various non-trivial checks.
However, the analytic proofs of these equations are still missing except for the $\text{U}(N)_k\times \text{U}(N+M)_{-k}$ ABJ theory in the limit with $k,-iM\rightarrow \infty$ (so called ``dual 4d limit'' \cite{Bonelli:2016idi}).
In this special limit the relation between the grand partition function and $\mathfrak{q}$-$\text{PIII}_3$ equation reduces to the relation between the grand partition function of the $O(2)$ matrix model and the $\text{PIII}_3$ differential equation, which was proved explicitly in \cite{Tracy:1995aaa}.
It would be interesting if we can generalize this proof to the ABJ theory before taking the dual 4d limit as well as the other M2-brane matrix models.
For the $N_\text{f}$ matrix models, it might also be possible to prove the biliear equations from the combinatorial formula, although we are still not successful.
Once we find such a proof for a known bilinear relation, it will also provide a new hint for finding similar relations for other M2-brane matrix models with Fermi gas formalism or residue sum formula.

It would also be interesting to address the leading $1/N$ non-perturbative effect we have determined \eqref{dNfpmpmconjecture} from the holographic viewpoint.
See e.g.~\cite{Gautason:2023igo,Beccaria:2023ujc,Gautason:2024nru,Gautason:2025per} for the related analysis.
In particular, here is an interesting observation on our result.
It is known that the partition function of ${\cal N}=4$ $\text{U}(N)$ super Yang-Mills theory with an adjonit hypermultiplet and $N_\text{f}=\ell$ hypermultiplets on the squashed three sphere $S^3_b$ can be calculated, for $b=\sqrt{2n-1}$ ($n\in\mathbb{N}$) and the mass parameter of the adjoint hypermultiplet $m$ tuned with respect to $b$ as $m=\frac{\pi i(b^2-3)}{2b}$, by the $S^3$ partition function \eqref{ADHMgeneralNf} of the theory with $N_\text{f}=n\ell$ hypermultiplets \cite{Hatsuda:2016uqa,Kubo:2024qhq,Kubo:2025dot} as
\begin{align}
Z^{(\ell),S^3_b}_{y_\alpha=\eta_\alpha}\Bigl(\xi,\frac{\pi i(b^2-3)}{2b};N\Bigr)=
Z^{(n\ell)}_{y_\alpha=\tilde{\eta}_\alpha}\Bigl(\sqrt{b}\xi,\frac{\pi i}{b^2};N\Bigr),\quad (b=\sqrt{2n-1})
\label{S3bbyS3}
\end{align}
with
\begin{align}
\tilde{\eta}_{n(\beta-1)+\gamma}=\frac{\eta_\beta}{b}+\frac{2\pi}{b^2}\Bigl(\gamma-\frac{n+1}{2}\Bigr),\quad (1\le \beta\le N_\text{f},\quad 1\le\gamma\le n).
\end{align}
Interestingly, we find that the worldsheet instanton coefficients $d^{(N_\text{f})}_{\omega_{\text{WS}\pm -}}(\zeta,m;y_\alpha)$ \eqref{dNfpmpmconjecture} for $N_\text{f}=n\ell$ vanishes identically with the particular choice of $m$ and $y_\alpha$ on the right-hand side of \eqref{S3bbyS3}.
As a result, the exponent of the actual leading non-perturbative effect will be $\omega_{\text{WS}\pm +}$ rather than $\omega_{\text{WS}\pm -}$ (here $\pm$ depends on the sign of $i\xi$).
In particular, let us focus on the case $b=\sqrt{3}$, where $\frac{\pi(b^2-3)}{2b}=0$ and hence we can calculate the $S^3_{b=\sqrt{3}}$ partition function in the conformal limit\footnote{
In order to preserve the conformal symmetry we also have to set $\eta_\alpha=0$.
However, this does not affect our argument on the leading instanton exponent.
}
$\xi=m=0$ by the $S^3$ partition function as
\begin{align}
Z^{(\ell),S^3_{b=\sqrt{3}}}_{y_\alpha=\eta_\alpha}(0,0;N)=
Z^{(2\ell)}_{y_\alpha=\tilde{\eta}_\alpha}\Bigl(0,\frac{\pi i}{3};N\Bigr),
\end{align}
with
\begin{align}
\tilde{\eta}_{2\beta-1}=\eta_\beta-\frac{\pi i}{3},\quad 
\tilde{\eta}_{2\beta}=\eta_\beta+\frac{\pi i}{3},\quad (\beta=1,\cdots,\ell).
\end{align}
In this case, if we write down the leading non-perturbative effect in the free energy $-\log Z(N)$, we find
\begin{align}
&-\log Z_{\eta_\alpha}^{(\ell),S^3_{b=\sqrt{3}}}\Bigl(0,0;N\Bigr) -\Bigl(-\log Z_{\text{pert},\eta_\alpha}^{(\ell),S^3_{b=\sqrt{3}}}\Bigl(0,0;N\Bigr)\Bigr)\nonumber \\
&=-\log Z_{\tilde{\eta}_\alpha}^{(2\ell)}\Bigl(0,\frac{\pi i}{3};N\Bigr) -\Bigl(-\log Z_{\text{pert},\tilde{\eta}_\alpha}^{(2\ell)}\Bigl(0,\frac{\pi i}{3};N\Bigr)\Bigr)\nonumber \\
&\sim \text{exp}\biggl[-\Bigl[\omega_{\text{WS}\pm +}\Bigr]_{N_\text{f}=2\ell,\zeta=0,m=\frac{\pi i}{3}}\sqrt{\frac{N}{[C]_{N_\text{f}=2\ell,\zeta=0,m=\frac{\pi i}{3}}
}}\biggr]\nonumber \\
&=e^{-2\pi\sqrt{\frac{2N}{\ell}}}
\label{leadinginstexpS3bsqrt3}
\end{align}
Here $C$
is given by \eqref{CofCBA} and $[\omega_{\text{WS}\pm +}]_{N_\text{f}=2\ell,\zeta=0,m=\frac{\pi i}{3}}=\frac{3}{\ell}$ is the worldsheet instanton exponent \eqref{omegaWSpmpm} in $N_\text{f}=2\ell$ matrix model with $\zeta=0$ and $m=\frac{\pi i}{3}$.
Interestingly, this leading instanton exponent \eqref{leadinginstexpS3bsqrt3} for the partition function on the squashed sphere with $b=\sqrt{3}$ precisely coincides with the leading instanton exponent for the $N_\text{f}=\ell$ matrix model with $\zeta=m=0$, which is\footnote{
Note that each worldsheet instanton coefficient $d_{\omega_{\text{WS}\pm\pm'}}^{(\ell)}(\xi,m;\eta_\alpha)$ for the $S^3$ partition function diverges for $\xi=m=0$.
However, after summing over the four contributions we are left with a finite coefficient
\begin{align}
&\lim_{m\rightarrow 0}\sum_{\pm,\pm'}d^{(\ell)}_{\omega_{\text{WS}\pm\pm'}}(0,m;y_\alpha)
\text{exp}\biggl[-\Bigl[\omega_{\text{WS}\pm\pm'}\Bigr]_{N_\text{f}=\ell,\zeta=0}\mu\biggr]\nonumber \\
&=\frac{1}{2\pi \sin\frac{2\pi}{\ell}}\sum_{\alpha=1}^{\ell}\Bigl(
-\frac{4\mu\cosh(2\eta_\alpha)}{\ell}
+2\eta_\alpha \sinh 2\eta_\alpha
-\cosh 2\eta_\alpha
\Bigr)e^{-\frac{4\mu}{\ell}}.
\end{align}
Note that this coefficient remains non-vanishing even if we set $\eta_\alpha=0$.
Hence the worldsheet instanton exponents $\omega_{\text{WS}\pm\pm'}$ actually give the exponent of the leading non-perturbative effect for $\zeta=m=0$.
}
\begin{align}
&-\log Z_{\eta_\alpha}^{(\ell)}\Bigl(0,0;N\Bigr) -\Bigl(-\log Z_{\text{pert},\eta_\alpha}^{(\ell)}\Bigl(0,0;N\Bigr)\Bigr)\nonumber \\
&\sim \text{exp}\biggl[-\Bigl[\omega_{\text{WS}\pm\pm'}\Bigr]_{N_\text{f}=\ell,\zeta=m=0}\sqrt{\frac{N}{[C]_{N_\text{f}=\ell,\zeta=m=0}}}\biggr]\nonumber \\
&=e^{-2\pi\sqrt{\frac{2N}{\ell}}}.
\end{align}
This observation may suggest that the exponent of the leading non-perturbative effect in $1/N$ is independent of the squashing parameter $b$ when the model is conformal, which is also mentioned in \cite{Gautason:2025per}.
It would be an interesting future work to examine this property for other values of $b$ as well as in different theories of M2-branes.

\acknowledgments

We are grateful to
Mykola Dedushenko,
Amihay Hanany,
Junho Hong,
Minxin Huang,
Marcos Mari\~no,
Satoshi Nawata,
Jun Nian,
Tadashi Okazaki,
Yuezhang Tang,
Xin Wang,
Congkao Wen,
Leopoldo A.~Pando Zayas,
Xuao Zhang and
Wenni Zheng
for valuable discussions.
The work of T.N.~was supported by the Startup Funding no.~2302-SRFP-2024-0012 of Shanghai Institute for Mathematics and Interdisciplinary Sciences.
We thank APCTP, Pohang, Korea, for their hospitality during the External Program [APCTP-2025-E09], from which this work greatly benefited.

\appendix

\section{Notation for Young diagram}
\label{app_Youngdiagramnotation}

For the integer partition $(\lambda_1,\lambda_2,\cdots,\lambda_n)$ of $N$ ($\lambda_1\le \lambda_2\le \cdots\le\lambda_n$, $\sum_{a=1}^n\lambda_a=N$), we associated a Young diagram $\lambda$ with $n$ rows, each of which contains $\lambda_i$ boxes.
We call $n$ as the length of the Young diagram and also denote it as $\ell(\lambda)$.
Given a Young diagram $\lambda=(\lambda_1,\cdots,\lambda_{\ell(\lambda)})$, we also define its transpose $\tilde{\lambda}$ as
\begin{align}
\tilde{\lambda}=(\tilde{\lambda}_1,\cdots,\tilde{\lambda}_{\ell(\tilde{\lambda})})
\label{transposeoflambda}
\end{align}
with
\begin{align}
\tilde{\lambda}_a=\#\{i|\lambda_i\ge a\},\quad (a=1,2,\cdots,\ell(\tilde{\lambda}))
\end{align}
and $\ell(\tilde{\lambda})=\lambda_1$.

\begin{figure}
\begin{center}
\includegraphics[width=8cm]{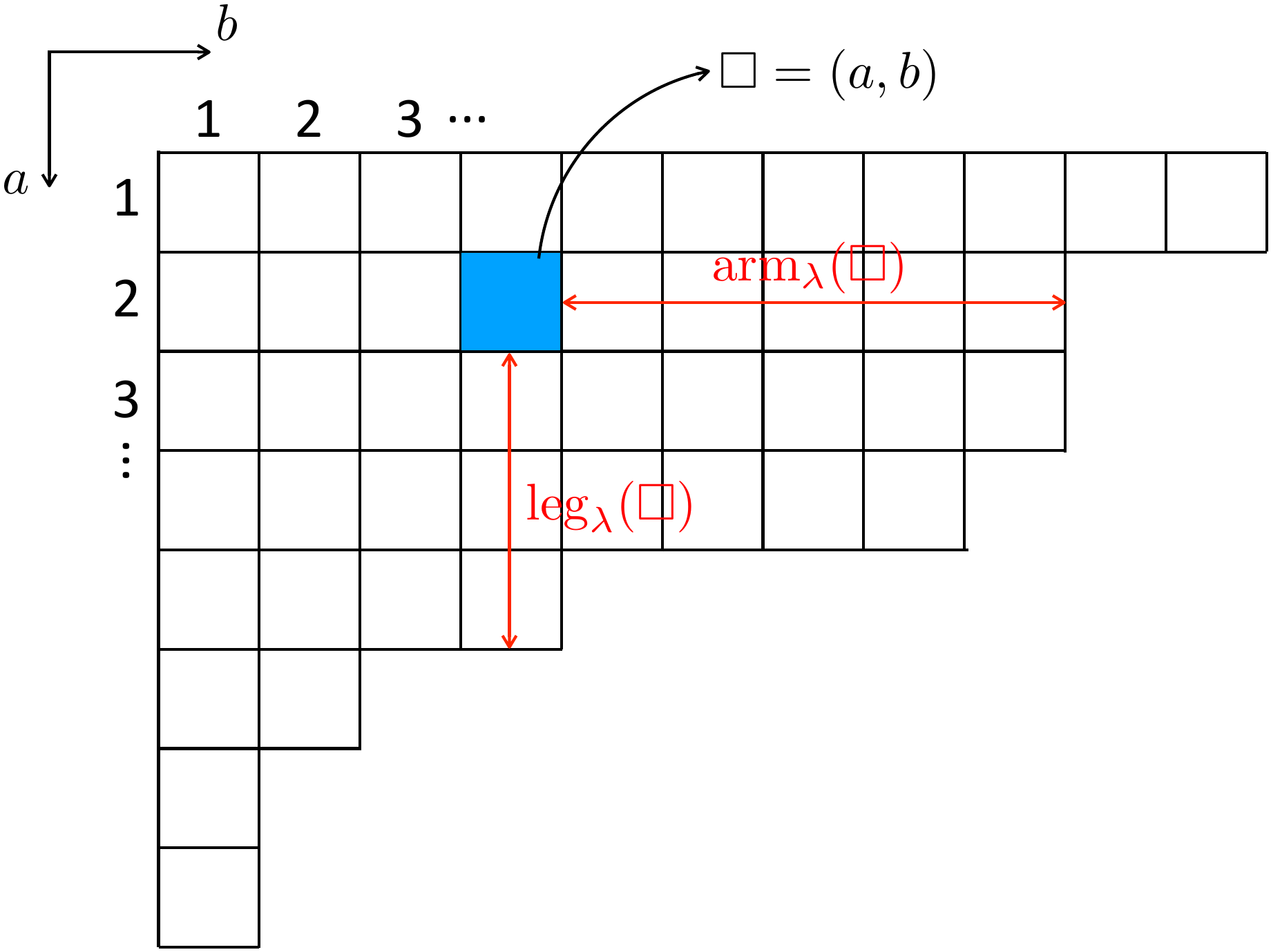}
\caption{
Graphical definition of arm/leg length \eqref{armleg1} of a box in a Young diagram.
}
\label{armlegconvention}
\end{center}
\end{figure}

We denote the box at the intersection of the $a$-th row and the $b$-th column as $\Box=(a,b)$.
For each $\Box=(a,b)\in\lambda$, we define the arm length $\text{arm}_\lambda(\Box)$ and the leg length $\text{leg}_\lambda(\Box)$ as
\begin{align}
\text{arm}_\lambda(\Box)=\lambda_a-b,\quad
\text{leg}_\lambda(\Box)=\tilde{\lambda}_b-a,
\quad (\Box=(a,b)\in\lambda).
\label{armleg1}
\end{align}
See figure \ref{armlegconvention}.
We also define the arm/leg length for a box outside of the Young diagram as
\begin{align}
\text{arm}_\lambda(\Box)=\begin{cases}
\lambda_a-b&\quad (a\le \ell(\lambda))\\
-b&\quad (a>\ell(\lambda))
\end{cases},\quad
\text{leg}_\lambda(\Box)=\begin{cases}
\tilde{\lambda}_b-a&\quad (b\le \lambda_1)\\
-a&\quad (b>\lambda_1)
\end{cases}.
\label{armleg2}
\end{align}

For each $\Box=(a,b)\in\lambda$, we define the hook length $h_\lambda(\Box)$ as
\begin{align}
h_\lambda(\Box)=\lambda_a-b+\tilde{\lambda}_b-a+1
=\text{arm}_\lambda(\Box)
+\text{leg}_\lambda(\Box)+1.
\label{hook}
\end{align}

\section{Proof of \eqref{gtildevsNekrasov}}
\label{app_proofofgtildevsNekrasov}

In this appendix we want to show the identity \eqref{gtildevsNekrasov}
\begin{align}
g_{\lambda,\mu}(y)
=
E_{\lambda,\mu}(y)
\times
\prod_{\Box\in\lambda}e^{-\frac{\pi i(\text{arm}_\lambda(\Box)-\text{leg}_\mu(\Box)}{2}}
\prod_{\Box\in\mu}e^{\frac{\pi i(\text{arm}_\mu(\Box)-\text{leg}_\lambda(\Box)}{2}}
\label{gtildevsNekrasovapp}
\end{align}
where
\begin{align}
&E_{\lambda,\mu}(y)=
\frac{1}{\displaystyle \prod_{\Box\in\lambda}
\Bigl(e^{\frac{y}{2}+\frac{m}{2}(\text{arm}_\lambda(\Box)+\text{leg}_\mu(\Box)+1)}+e^{-\frac{y}{2}-\frac{m}{2}(\text{arm}_\lambda(\Box)+\text{leg}_\mu(\Box)+1)-\pi i(\text{arm}_\lambda(\Box)-\text{leg}_\mu(\Box))}\Bigr)
}\nonumber \\
&\quad\quad\times \frac{1}{\displaystyle \prod_{\Box\in\mu}
\Bigl(e^{\frac{y}{2}-\frac{m}{2}(\text{arm}_\mu(\Box)+\text{leg}_\lambda(\Box)+1)}+e^{-\frac{y}{2}+\frac{m}{2}(\text{arm}_\mu(\Box)+\text{leg}_\lambda(\Box)+1)+\pi i(\text{arm}_\mu(\Box)-\text{leg}_\lambda(\Box))}\Bigr)
},
\end{align}
and
\begin{align}
&g_{\lambda,\mu}(y)
=
\left(
\prod_{\Box=(a,b)\in\lambda}\prod_{\Box'=(a',b')\in\mu}
\frac{
(2\sinh\frac{y+(b-b'-a+a')m+\pi i(a+b-a'-b')}{2})^2
}{
\prod_\pm 2\cosh\frac{y+(b-b'-a+a'\pm 1)m+\pi i(a+b-a'-b')}{2}
}
\right)\nonumber \\
&\quad\times \left(
\prod_{\Box=(a,b)\in\lambda}
\frac{1}{2\cosh\frac{y+(b-a)m+\pi i (a+b-1)}{2}}
\right)
\left(
\prod_{\Box=(a,b)\in\mu}
\frac{1}{2\cosh\frac{-y+(b-a)m+\pi i(a+b-1)}{2}}
\right).
\label{g}
\end{align}
In order to prove the identity \eqref{gtildevsNekrasovapp}, let us first define the following index
\begin{align}
I^{(g)}_{\lambda,\mu}(\xi,\eta)=&\sum_{a=1}^{\tilde{\lambda}_1}\sum_{b=1}^{\lambda_a}\xi^b\eta^a
+\sum_{b=1}^{\mu_1}\sum_{a=1}^{\tilde{\mu}_b}\xi^{1-b}\eta^{1-a}\nonumber\\
&+\sum_{\Box=(a,b)\in\lambda}\sum_{\Box'=(a',b')\in\mu}\left(\xi^{b-b'+1}-\xi^{b-b'}\right)\left(\eta^{a-a'+1}-\eta^{a-a'}\right),
\label{index}
\end{align}
which satisfies the following symmetry property
\begin{align}
I^{(g)}_{\lambda,\mu}(\xi,\eta)=I^{(g)}_{\tilde{\lambda},\tilde{\mu}}(\eta,\xi).
\label{symmetryofI_notproved}
\end{align}
Here $\tilde{\lambda}$ and $\tilde{\mu}$ stand for the transpose of the Young diagrams $\lambda$ and $\mu$ defined as \eqref{transposeoflambda}.
Given any two-variable function $f(a,b)$, the index \eqref{index} encodes a function $F$ given in a product expression of $f(a,b)$ as
\begin{align}
I^{(g)}_{\mu,\nu}(\xi,\eta)=\sum_{(\alpha,a,b)}\alpha \xi^a \eta^b \quad \longrightarrow \quad F=\prod_{(\alpha,a,b)}\frac{1}{f(a,b)^\alpha},
\label{ItoF}
\end{align}
or explicitly
\begin{align}
&I^{(g)}_{\mu,\nu}(\xi,\eta)\rightarrow F=\left(
\prod_{a=1}^{\tilde{\lambda}_1}\prod_{b=1}^{\lambda_a}\frac{1}{f(b,a)}
\right)
\left(
\prod_{b=1}^{\mu_1}\prod_{a=1}^{\tilde{\mu}_b}\frac{1}{f(1-b,1-a)}
\right)\nonumber\\
&\quad \quad \quad \times \left(
\prod_{\Box=(a,b)\in\lambda}
\prod_{\Box'=(a',b')\in\mu}
\frac{f(b-b'+1,a-a'+1)f(b-b',a-a')}{f(b-b'+1,a-a')f(b-b',a-a'+1)}
\right).
\label{product}
\end{align}
By choosing the function $f(a,b)$ as 
\begin{align}
f(a,b) = 2 \cosh \frac{y+(a-b)m + \pi i (a+b -1)}{2} \nonumber
\label{choosef}
\end{align}
the product \eqref{product} coincides with $g_{\lambda, \mu}(y)$ defined in \eqref{g}. 
Similarly, after rewriting the right-hand side of \eqref{gtildevsNekrasovapp} as
\begin{align}
&E_{\lambda,\mu}(y)
\times
\prod_{\Box\in\lambda}e^{-\frac{\pi i(\text{arm}_\lambda(\Box)-\text{leg}_\mu(\Box)}{2}}
\prod_{\Box\in\mu}e^{\frac{\pi i(\text{arm}_\mu(\Box)-\text{leg}_\lambda(\Box)}{2}}\nonumber \\
&=
\prod_{\Box\in\lambda}\frac{1}{2 \cosh \frac{y+(\text{arm}_\lambda(\Box) + \text{leg}_\mu(\Box) + 1)m + \pi i (\text{arm}_\lambda(\Box) - \text{leg}_\mu(\Box))}{2}}\nonumber \\
&\quad \times \prod_{\Box\in\mu}\frac{1}{2 \cosh \frac{y+(-\text{arm}_\mu(\Box) - \text{leg}_\lambda(\Box) - 1)m + \pi i (-\text{arm}_\mu(\Box) + \text{leg}_\lambda(\Box))}{2}},
\end{align}
we find that it is encoded by the following index $I^{(E)}_{\lambda,\mu}(\xi,\eta)$ through the same rule \eqref{ItoF} and the same choice of $f(a,b)$ \eqref{choosef}
\begin{align}
I^{(E)}_{\lambda,\mu}(\xi,\eta)=
\sum_{\Box\in\lambda}\xi^{\text{arm}_\lambda(\Box)+1}\eta^{-\text{leg}_\mu(\Box)}
+\sum_{\Box\in\mu}\xi^{-\text{arm}_\mu(\Box)}\eta^{\text{leg}_\lambda(\Box)+1}.
\label{IE}
\end{align}

Therefore, the identity \eqref{gtildevsNekrasovapp} can be proved by proving the identity of the indices, $I^{(g)}_{\lambda,\mu}(\xi,\eta)=I^{(E)}_{\lambda,\mu}(\xi,\eta)$.
For this purpose, 
we first rewrite the third part of $I^{(g)}_{\lambda,\mu}(\xi,\eta)$ as
\begin{align}
\sum_{\Box=(i,a)\in\lambda}\sum_{\Box'=(b,j)\in\mu} &\left(\xi^{a+1-j}-\xi^{a-j}\right)\left(\eta^{i-b}-\eta^{i+1-b}\right)\nonumber \\
&=\sum_{i=1}^{\tilde{\lambda}_1}\sum_{a=1}^{\lambda_i}\sum_{j=1}^{\mu_1}\sum_{b=1}^{\tilde{\mu}_j}\left(\xi^{a+1-j}-\xi^{a-j}\right)\left(\eta^{i-b}-\eta^{i+1-b}\right),
\label{sum}
\end{align}
then we rewrite the sum over $a$ into 
\begin{align}
\sum_{a=1}^{\lambda_i}(\xi^{a+1-j}-\xi^{a-j})=\sum_{a=2}^{\lambda_i+1}\xi^{a-j}-\sum_{a=1}^{\lambda_i}\xi^{a-j}=\xi^{\lambda_i-j+1}-\xi^{-j+1}.
\end{align}
Similarly, we can rewrite the sum over $b$ into 
\begin{align}
\sum_{b=1}^{\tilde{\mu}_j}(\eta^{i-b}-\eta^{i-b+1})=\eta^{-\tilde{\mu}_j+i}-\eta^i.
\end{align}
Then we can write \eqref{sum} as
\begin{align}
\sum_{\Box=(i,a)\in\lambda}\sum_{\Box'=(b,j)\in\mu}(\xi^{a+1-j}-\xi^{a-j})(\eta^{i-b}-\eta^{i+1-b})=\sum_{i=1}^{\tilde{\lambda}_1}\sum_{j=1}^{\mu_1}(\xi^{\lambda_i-j+1}-\xi^{-j+1})(\eta^{-\tilde{\mu}_j+i}-\eta^i),
\end{align}
and the index \eqref{index} is rewritten as 
\begin{align}
I^{(g)}_{\lambda,\mu}(\xi,\eta)=\sum_{a=1}^{\tilde{\lambda}_1}\sum_{b=1}^{\lambda_a}\xi^b\eta^a
+\sum_{b=1}^{\mu_1}\sum_{a=1}^{\tilde{\mu}_b}\xi^{1-b}\eta^{1-a}
+\sum_{a=1}^{\tilde{\lambda}_1}\sum_{b=1}^{\mu_1}\left(\xi^{\lambda_a-b+1}-\xi^{-b+1}\right)\left(\eta^{-\tilde{\mu}_b+a}-\eta^a\right).
\label{new index}
\end{align}

Next we expand the third term of $I^{(g)}_{\lambda,\mu}(\xi,\eta)$ \eqref{new index} and then add $0=\xi^{-b+1}\eta^a-\xi^{-b+1}\eta^a$ to the summand in the third term
\begin{align}
I^{(g)}_{\lambda,\mu}(\xi,\eta) =&\sum_{a=1}^{\tilde{\lambda}_1}\sum_{b=1}^{\lambda_a}\xi^b\eta^a
+\sum_{b=1}^{\mu_1}\sum_{a=1}^{\tilde{\mu}_b}\xi^{1-b}\eta^{1-a}
+\sum_{a=1}^{\tilde{\lambda}_1}\sum_{b=1}^{\mu_1}
\big(
\xi^{\lambda_a-b+1}\eta^{-\tilde{\mu}_b+a}
\color{blue}{
-\,\xi^{\lambda_a-b+1}\eta^a
}\nonumber\\
&\color{red}{
-\,\xi^{-b+1}\eta^{-\tilde{\mu}_b+a}
+\xi^{-b+1}\eta^a
}
 \color{blue}{+\,\xi^{-b+1}\eta^a}
\color{black}{ -\,\xi^{-b+1}\eta^a
\big).}
\end{align}
The summation over red and blue parts can be  reorganized as 
\begin{align}
\textcolor{red}{\text{red part}}&=\sum_{a=1}^{\tilde{\lambda}_1}\sum_{b=1}^{\mu_1}(-\xi^{-b+1})(\eta^{-\tilde{\mu}_b+a}-\eta^a)=\sum_{b=1}^{\mu_1}\sum_{a=1}^{\tilde{\mu}_b}(-\xi^{-b+1})(\eta^{1-a}-\eta^{\tilde{\lambda}_1-a+1}),
\end{align}
\begin{align}
\textcolor{blue}{\text{blue part}}&=\sum_{a=1}^{\tilde{\lambda}_1}\sum_{b=1}^{\mu_1}(-\eta^{a})(\xi^{\lambda_a-b+1}-\xi^{-b+1})=\sum_{a=1}^{\tilde{\lambda}_1}\sum_{b=1}^{\lambda_a}(-\eta^a)(\xi^b-\xi^{-\mu_1+b}).
\end{align}
Taken these into account, $I_{\lambda,\mu}(\xi,\eta)$ in \eqref{new index} can be rewritten as
\begin{align}
I^{(g)}_{\lambda,\mu}(\xi,\eta)=\sum_{a=1}^{\tilde{\lambda}_1}\sum_{b=1}^{\mu_1}(\xi^{\lambda_a-b+1}\eta^{-\tilde{\mu}_b+a}-\xi^{-b+1}\eta^a)
+\sum_{b=1}^{\mu_1}\sum_{a=1}^{\tilde{\mu}_b}\xi^{-b+1}\eta^{\tilde{\lambda}_1-a+1}
+\sum_{a=1}^{\tilde{\lambda}_1}\sum_{b=1}^{\lambda_a}\xi^{-\mu_1+b}\eta^{a}.
\label{Igsecond}
\end{align}

In order to obtain the expression $I^{(E)}_{\lambda,\mu}(\xi,\eta)$ \eqref{IE} from this expression of $I^{(g)}_{\lambda,\mu}(\xi,\eta)$ \eqref{Igsecond}, let us decompose $I^{(g)}_{\lambda,\mu}(\xi,\eta)$ into the positive powers of $\xi$, which we denote $I^{(+)}_{\lambda,\mu}(\xi,\eta)$, and the non-positive powers of $\xi$, which we denote $I^{(-)}_{\lambda,\mu}(\xi,\eta)$, as
\begin{align}
I^{(g)}_{\lambda,\mu}(\xi,\eta)=I_{\lambda,\mu}^{(+)}(\xi,\eta)+I_{\lambda,\mu}^{(0,-)}(\xi,\eta)
\label{IinI+andI0-}
\end{align}
and calculate $I^{(+)}$ and $I^{(0,-)}$ separately \cite{Nakajimabook}.
For $I_{\lambda,\mu}^{(+)}(\xi,\eta)$ we find that only part of the first term in the first summation and part of the third summation contribute, and we obtain
\begin{align}
I_{\lambda,\mu}^{(+)}(\xi,\eta)=\sum_{a=1}^{\tilde{\lambda}_1}\sum_{b=1}^{\text{min}(\mu_1,\lambda_a)}\xi^{\lambda_a-b+1}\eta^{-\tilde{\mu}_b+a}
+\sum_{a=1}^{\tilde{\lambda}_1}\sum_{b=\mu_1+1}^{\lambda_a}\xi^{-\mu_1+b}\eta^{a}.
\end{align}
Note that the summation over $b$ in the second term vanishes when $\mu_1+1>\lambda_a$.
Taking into account of this, we can rewrite the summation over $b$ in the second term as
\begin{align}
\sum_{b=\mu_1+1}^{\lambda_a}\xi^{-\mu_1+b}
= \!\!\!\!\!\!\!\!\! \sum_{b= \text{min}(\mu_1,\lambda_a)+1}^{\lambda_a} \!\!\!\!\!\!\!\!\! \xi^{-\text{min}(\mu_1,\lambda_a)+b}
=\xi+\xi^2+\cdots+\xi^{\lambda_a-\text{min}(\mu_1,\lambda_a)}
=\!\!\!\!\!\!\!\!\! \sum_{b=\text{min}(\mu_1,\lambda_a)+1}^{\lambda_a} \!\!\!\!\!\!\!\!\! \xi^{\lambda_a-b+1}.
\end{align}
Also note that since $b$ in this summation always satisfies $b>\mu_1$ (this is true for $\mu_1<\lambda_a$; when $\mu_1\ge \lambda_a$ the summation itself does not exist).
If we define $\tilde{\mu}_b=0$ for $b>\mu_1$, $I^{(+)}_{\lambda,\mu}(\xi,\eta)$ can be written as
\begin{align}
I_{\lambda,\mu}^{(+)}(\xi,\eta)&=\sum_{a=1}^{\tilde{\lambda}_1}\sum_{b=1}^{\text{min}(\mu_1,\lambda_a)}\xi^{\lambda_a-b+1}\eta^{-\tilde{\mu}_b+a}
+\sum_{a=1}^{\tilde{\lambda}_1}\sum_{b=\text{min}(\mu_1,\lambda_a)+1}^{\lambda_a}\xi^{\lambda_a-b+1}\eta^{-\tilde{\mu}_b+a}\nonumber \\
&=\sum_{a=1}^{\tilde{\lambda}_1}\sum_{b=1}^{\lambda_a}\xi^{\lambda_a-b+1}\eta^{-\tilde{\mu}_b+a}.
\label{I+final}
\end{align}
Here in the last expression we have just combined the two summations.

Next let us consider $I^{(0,-)}_{\lambda,\mu}(\xi,\eta)$.
For this purpose, now we can use the identity \eqref{symmetryofI_notproved} to write $I_{\lambda,\mu}(\xi,\eta)$ as
\begin{align}
I^{(g)}_{\lambda,\mu}(\xi,\eta)=\sum_{b=1}^{\lambda_1}\sum_{a=1}^{\tilde{\mu}_1}(\eta^{\tilde{\lambda}_b-a+1}\xi^{-\mu_a+b}-\eta^{-a+1}\xi^b)+\sum_{a=1}^{\tilde{\mu}_1}\sum_{b=1}^{\mu_a}\eta^{-a+1}\xi^{\lambda_1-b+1}+\sum_{b=1}^{\lambda_1}\sum_{a=1}^{\tilde{\lambda}_b}\eta^{-\tilde{\mu}_1+a}\xi^b.
\end{align}
In this expression only part of the first term in the first summation and part of the second summation contribute to $I^{(0,-)}_{\lambda,\mu}(\xi,\eta)$ as
\begin{align}
I^{(0,-)}_{\lambda,\mu}(\xi,\eta)
=
\sum_{a=1}^{\tilde{\mu}_1}
\sum_{b=1}^{\text{min}(\lambda_1,\mu_a)}\eta^{\tilde{\lambda}_b-a+1}\xi^{-\mu_a+b}
+\sum_{a=1}^{\tilde{\mu}_1}\sum_{b=\lambda_1+1}^{\mu_a}\eta^{-a+1}\xi^{\lambda_1-b+1}.
\end{align}
By doing similar calculation as we did for $I^{(+)}_{\lambda,\mu}(\xi,\eta)$, we obtain
\begin{align}
I^{(0,-)}_{\lambda,\mu}(\xi,\eta)=\sum_{a=1}^{\tilde{\mu}_1}\sum_{b=1}^{\mu_a}\xi^{-\mu_a+b}\eta^{\tilde{\lambda}_b-a+1}.
\label{I0-final}
\end{align}
Plugging \eqref{I+final} and \eqref{I0-final} into \eqref{IinI+andI0-}, we obtain
\begin{align}
I^{(g)}_{\lambda,\mu}(\xi,\eta)&=\sum_{a=1}^{\tilde{\lambda}_1}\sum_{b=1}^{\lambda_a}\xi^{\lambda_a-b+1}\eta^{-\tilde{\mu}_b+a} + \sum_{a=1}^{\tilde{\mu}_1}\sum_{b=1}^{\mu_a}\xi^{-\mu_a+b}\eta^{\tilde{\lambda}_b-a+1}\nonumber\\
&=\sum_{\Box\in\lambda}\xi^{\text{arm}_\lambda(\Box)+1}\eta^{-\text{leg}_\mu(\Box)}
+\sum_{\Box\in\mu}\xi^{-\text{arm}_\mu(\Box)}\eta^{\text{leg}_\lambda(\Box)+1},
\end{align}
which coincides with $I^{(E)}_{\lambda,\mu}(\xi,\eta)$ \eqref{IE}.
As mentioned below \eqref{IE}, this completes the proof of the identity \eqref{gtildevsNekrasovapp}.

\bibliographystyle{utphys}
\bibliography{references.bib}

\end{document}